\documentclass[twocolumn,9pt]{article}

\pdfoutput=1
\usepackage{extsizes}
\usepackage[super,sort&compress,comma]{natbib} 
\usepackage[version=3]{mhchem}
\usepackage[left=1.5cm, right=1.5cm, top=1.785cm, bottom=2.0cm]{geometry}
\usepackage{balance}
\usepackage{widetext}
\usepackage{times,mathptmx}
\usepackage{sectsty}
\usepackage{graphicx} 
\usepackage{lastpage}
\usepackage[format=plain,justification=raggedright,singlelinecheck=false,font={stretch=1.125,small,sf},labelfont=bf,labelsep=space]{caption}
\usepackage{float}
\usepackage{fancyhdr}
\usepackage{fnpos}
\usepackage[english]{babel}
\usepackage{array}
\usepackage{droidsans}
\usepackage{charter}
\usepackage[T1]{fontenc}
\usepackage[usenames,dvipsnames]{xcolor}
\usepackage{setspace}
\usepackage[compact]{titlesec}

\usepackage{epstopdf}

\definecolor{cream}{RGB}{222,217,201}

\newcommand*{\citen}[1]{%
  \begingroup
    \romannumeral-`\x 
    \setcitestyle{numbers}%
    \cite{#1}%
  \endgroup   
}

\begin{document}

\pagestyle{fancy}
\thispagestyle{plain}
\fancypagestyle{plain}{

\fancyhead[C]{\includegraphics[width=18.5cm]{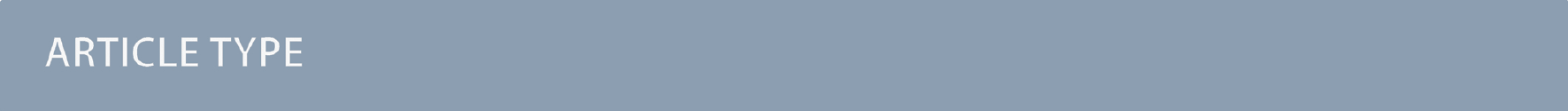}}
\fancyhead[L]{\hspace{0cm}\vspace{1.5cm}\includegraphics[height=30pt]{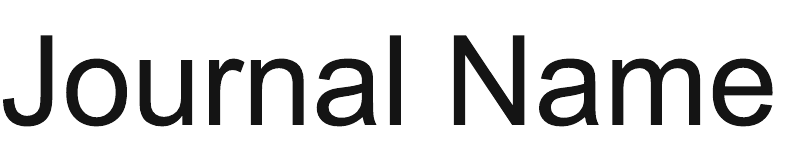}}
\fancyhead[R]{\hspace{0cm}\vspace{1.7cm}\includegraphics[height=55pt]{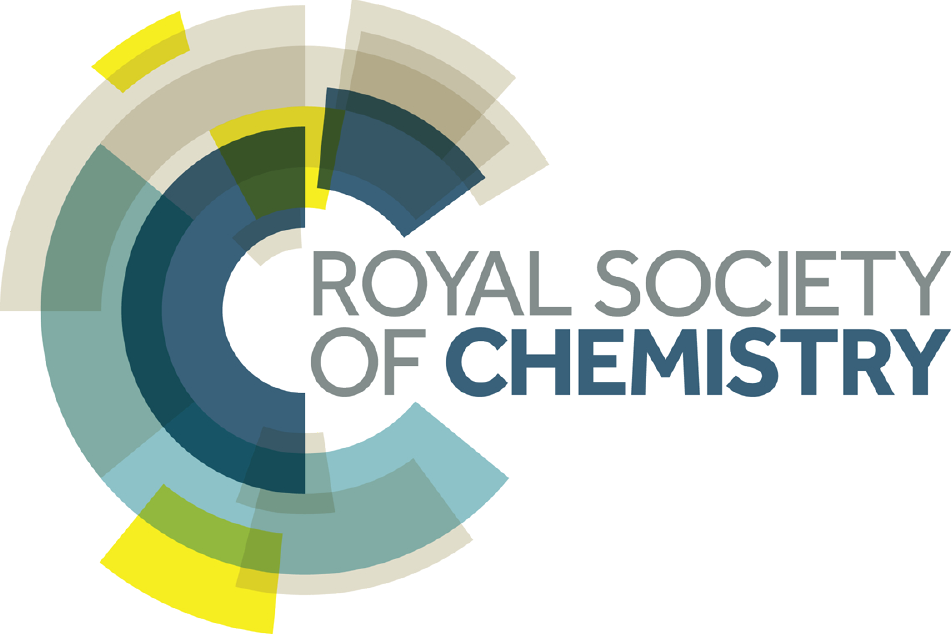}}
\renewcommand{\headrulewidth}{0pt}
}

\makeFNbottom
\makeatletter
\renewcommand\LARGE{\@setfontsize\LARGE{15pt}{17}}
\renewcommand\Large{\@setfontsize\Large{12pt}{14}}
\renewcommand\large{\@setfontsize\large{10pt}{12}}
\renewcommand\footnotesize{\@setfontsize\footnotesize{7pt}{10}}
\makeatother

\renewcommand{\thefootnote}{\fnsymbol{footnote}}
\renewcommand\footnoterule{\vspace*{1pt}%
\color{cream}\hrule width 3.5in height 0.4pt \color{black}\vspace*{5pt}} 
\setcounter{secnumdepth}{5}

\makeatletter 
\renewcommand\@biblabel[1]{#1}            
\renewcommand\@makefntext[1]%
{\noindent\makebox[0pt][r]{\@thefnmark\,}#1}
\makeatother 
\renewcommand{\figurename}{\small{Fig.}~}
\sectionfont{\sffamily\Large}
\subsectionfont{\normalsize}
\subsubsectionfont{\bf}
\setstretch{1.125} 
\setlength{\skip\footins}{0.8cm}
\setlength{\footnotesep}{0.25cm}
\setlength{\jot}{10pt}
\titlespacing*{\section}{0pt}{4pt}{4pt}
\titlespacing*{\subsection}{0pt}{15pt}{1pt}

\fancyfoot{}
\fancyfoot[LO,RE]{\vspace{-7.1pt}\includegraphics[height=9pt]{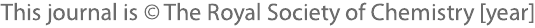}}
\fancyfoot[CO]{\vspace{-7.1pt}\hspace{13.2cm}\includegraphics{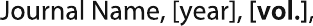}}
\fancyfoot[CE]{\vspace{-7.2pt}\hspace{-14.2cm}\includegraphics{RF}}
\fancyfoot[RO]{\footnotesize{\sffamily{1--\pageref{LastPage} ~\textbar  \hspace{2pt}\thepage}}}
\fancyfoot[LE]{\footnotesize{\sffamily{\thepage~\textbar\hspace{3.45cm} 1--\pageref{LastPage}}}}
\fancyhead{}
\renewcommand{\headrulewidth}{0pt} 
\renewcommand{\footrulewidth}{0pt}
\setlength{\arrayrulewidth}{1pt}
\setlength{\columnsep}{6.5mm}
\setlength\bibsep{1pt}

\makeatletter 
\newlength{\figrulesep} 
\setlength{\figrulesep}{0.5\textfloatsep} 

\newcommand{\topfigrule}{\vspace*{-1pt}%
\noindent{\color{cream}\rule[-\figrulesep]{\columnwidth}{1.5pt}} }

\newcommand{\botfigrule}{\vspace*{-2pt}%
\noindent{\color{cream}\rule[\figrulesep]{\columnwidth}{1.5pt}} }

\newcommand{\dblfigrule}{\vspace*{-1pt}%
\noindent{\color{cream}\rule[-\figrulesep]{\textwidth}{1.5pt}} }

\makeatother

\twocolumn[
  \begin{@twocolumnfalse}
\vspace{3cm}
\sffamily
\begin{tabular}{m{4.5cm} p{13.5cm} }

\includegraphics{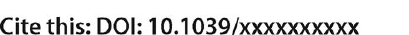} & 
\noindent\LARGE{\textbf{Segregation of polymers under cylindrical confinement: Effects of 
polymer topology and crowding}} 
\\
\vspace{0.3cm} & \vspace{0.3cm} \\

 & \noindent\large{James M. Polson,$^{\ast}$ and Deanna R.-M. Kerry\textit{$^{\ddag}$}} 
\\

\includegraphics{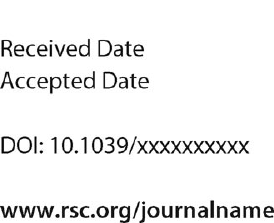} & 
\noindent\normalsize{%
Monte Carlo computer simulations are used to study the segregation behaviour of two 
polymers under cylindrical confinement. Using a multiple-histogram method, the 
conformational free energy, $F$, of the polymers was measured as a function of 
the centre-of-mass separation distance, $\lambda$. We examined the scaling of the 
free energy functions with the polymer length, the length and diameter of the 
confining cylinder, the polymer topology (i.e.  linear vs ring polymers), and 
the packing fraction and size of mobile crowding agents. In the absence of
crowders, the observed scaling of $F(\lambda)$ is similar to that predicted using 
a simple model employing the de~Gennes blob model and the approximation that the 
free energy of overlapping chains in a tube is equal to that of two isolated
chains each in a tube of half the cross-sectional area. Simulations were used to test 
the latter approximation and reveal that it yields poor quantitative predictions. 
This, along with generic finite-size effects, likely gives rise to the discrepancies 
between the predicted and measured values of scaling exponents for $F(\lambda)$. 
For segregation in the presence of crowding agents, the free energy barrier
generally decreases with increasing crowder packing fraction, thus reducing the entropic
forces driving segregation. However, for fixed packing fraction, the barrier 
increases as the crowder/monomer size ratio decreases.
} 

\end{tabular}

 \end{@twocolumnfalse} \vspace{0.6cm}

]


\footnotetext{\textit{Department of Physics, University of Prince Edward Island, 550 University Ave.,
Charlottetown, Prince Edward Island, C1A 4P3, Canada.  E-mail: jpolson@upei.ca}}


\footnotetext{\ddag~ Present address: Department of Physics and Atmospheric Science,
Dalhousie University, Halifax, Nova Scotia, B3H 4R2, Canada.}


\section{Introduction}
\label{sec:intro}

Overlapping polymers confined to a narrow channel experience a reduction in their
conformational entropy, leading to an effective repulsion that drives separation of the 
chains.\cite{daoud1977statistics,ha2015polymers}  This effect has been the subject of 
much study in recent years due to its possible role as a mechanism for chromosome 
separation in replicating bacteria.\cite{jun2006entropy,jun2010entropy,youngren2014multifork}
While recent evidence suggests that entropic forces alone may be insufficient to 
achieve complete segregation of chromosomes,\cite{lechat2012let,yazdi2012variation,%
kuwada2013mapping,diventura2013chromosome,junier2014polymer,lampo2015physical}
a complete picture of the mechanism of bacterial chromosome segregation 
continues to be elusive.\cite{reyes2012chromosome,wang2013organization,diventura2013chromosome,%
badrinarayanan2015bacterial,hajduk2016connecting} It remains plausible that
entropy makes a significant contribution to this process. Thus, using the basic theoretical
tools of polymer physics to study separation of confined polymers should help 
elucidate its role in bacterial chromosome segregation.\cite{ha2015polymers}
In addition, such theoretical insight will be of value for interpreting results
of in vitro experiments of DNA separation in nanochannels.\cite{liu2018probing}

Polymer segregation under confinement in channels has been the subject of numerous simulation 
studies\cite{jun2006entropy,teraoka2004computer,jun2007confined,arnold2007time,%
jacobsen2010demixing,jung2010overlapping,jung2012ring,jung2012intrachain,%
liu2012segregation,dorier2013modelling,racko2013segregation,shin2014mixing,%
minina2014induction,minina2015entropic,chen2015polymer,polson2014polymer} 
These studies have mostly employed simple bead-spring polymer models composed 
of ${\cal O}(10^2)$
monomers. In relation to the bacterial chromosome, it has been proposed that each bead can 
be viewed as a representation of a topological domain of ${\cal O}(10^4$--$10^5)$ base pairs%
\cite{pelletier2012physical} formed by a combination of negative supercoiling of the DNA 
and stabilization by nucleoid-assisted proteins.\cite{ha2015polymers} Generally, the focus 
of these simulations has been characterizing the demixing dynamics of initially 
overlapping chains, as well as quantifying the degree of miscibility of separated 
polymers in equilibrium.  Most have examined fully flexible 
linear polymers,\cite{jun2006entropy,teraoka2004computer,jun2007confined,arnold2007time,%
jacobsen2010demixing,jung2010overlapping,jung2012intrachain,liu2012segregation,%
racko2013segregation,minina2014induction,polson2014polymer} though a number have also 
considered ring polymers.\cite{jung2012ring,dorier2013modelling,shin2014mixing,%
minina2015entropic,chen2015polymer}
Others have considered the effects of of bending rigidity\cite{racko2013segregation,%
polson2014polymer} and macromolecular crowding.\cite{shin2014mixing,chen2015polymer}
Typically, the results are analyzed and interpreted using analytical approximations
for the variation of the conformational free energy with the separation 
distance of the polymer centres of mass. For example, Refs.~\citen{racko2013segregation} and 
\citen{minina2015entropic} employ an expression for the free energy derived using the 
de~Gennes blob model in combination with approximating the free energy of two overlapping 
chains in a tube to be equal to that of two polymers in separate tubes, each with half the 
cross-sectional area of the real channel.\cite{jung2012ring} 
While such approximations are convenient and conceptually
elegant, finite-size effects for the blob model are expected to be significant for the system 
sizes typically employed in such simulations.\cite{kim2013elasticity} In addition, the accuracy 
of the approximation suggested in Ref.~\citen{jung2012ring} has thus far not been measured. 
Consequently, some caution is required when using such expressions for the free energy for 
a quantitative analysis of polymer segregation dynamics.

An alternative to using such analytical approximations is to measure the free energy function 
directly in simulations by means of probability histograms. Surprisingly, there have 
been only a few such studies. For example, Shin {\it et al.} 
measured a relatively small overlap barrier for overlapping ring polymers subject to 
confinement in a tube of finite length,\cite{shin2014mixing} while Minina and Arnold 
calculated and characterized the portion of the free energy function associated with 
the initial induction phase of segregation.\cite{minina2014induction,minina2015entropic} 
One complication that arises in the calculation of free energy functions for overlapping
polymers under strong confinement is the presence of large overlap free energy barriers. 
In such cases, more advanced simulation methods can be required to avoid poor statistics.  
Recently, we used Monte Carlo (MC) simulations employing umbrella sampling with a multiple 
histogram method to calculate the overlap free energy functions of linear polymers for a hard-sphere 
chain model system.\cite{polson2014polymer} We studied the scaling of the functions with polymer 
length and confinement dimensions for both infinite- and finite-length tubes, as well as the 
effects of bending rigidity. Generally, we found the results to be in reasonable agreement with 
the analytical predictions, with quantitative discrepancies in scaling exponents that were thought
to be due to finite-size effects. One notable observation for flexible chains was the presence 
of a regime over a range of centre-of-mass separations in which the polymers where in contact 
and compressed, but not overlapping. MC dynamics calculations revealed that upon separation 
from an initially fully overlapping state, the polymers initially remained in conformational 
quasi-equilibrium until this regime was entered.\cite{polson2014polymer}

The purpose of this study is to continue our examination of the properties of the free energy
function for a system of two cylindrically confined polymers. We employ the same MC methods
used previously to study similar hard-sphere model systems. We focus on the effect of 
polymer topology by carrying out calculations for ring polymers and comparing the results 
with those of linear polymers. These calculations are relevant for chromosomes of bacteria 
such as {\it E. coli}, which possess ring topology. As in Ref.~\citen{polson2014polymer} we 
examine the scaling of the free energies with respect to polymer length and confinement 
dimensions and compare the results with predictions from a simple analytical model. 
We also carry out calculations to measure the accuracy of the approximation of 
Ref.~\citen{jung2012ring} employed in the analytical model and find a significant quantitative
discrepancy with the true overlap free energy. We also study the effect of macromolecular 
crowding on the free energy function by incorporating mobile crowding agents into the system. 
Crowding effects are likely important for bacteria, within which approximately 30--35\%
of the volume fraction is occupied by RNA, ribosomes and other biomacromolecules.
We examine the effect of varying both the packing fraction and the size of the crowding agents. 
Generally, we find that increasing the crowder density decreases the barrier height, while at 
fixed crowder packing fraction, decreasing the size of the crowding agents increases the barrier 
height.

This article is organized as follows. In Section~\ref{sec:model} we briefly describe the
model used in the study, while Section~\ref{sec:methods} outlines the MC method used to
calculate the free energy functions. Section~\ref{sec:results} presents the main results
of the study, which are interpreted and discussed in detail. Results for ring and linear
polymer systems in the absence of crowding agents are presented, for both infinite- and 
finite-length confining cylinders, after which results for systems with crowding agents 
are presented. In Section~\ref{sec:conclusions} we summarize the key findings of this
study.

\section{Model}
\label{sec:model}

We employ a minimal model of two polymer chains confined to a cylindrical tube. Each polymer is
modeled as a chain of $N$ hard spheres, each with a diameter of $\sigma$. Thus, the pair potential for
non-bonded monomers is $u_{\rm{nb}}(r)=\infty$ for $r\leq\sigma$ and $u_{\rm{nb}}(r)=0$ for
$r>\sigma$, where $r$ is the distance between the centres of the monomers. Pairs of bonded monomers
interact with a potential $u_{\rm{b}}(r)= 0$ if $0.9\sigma<r<1.1\sigma$ and $u_{\rm{b}}(r)= \infty$,
otherwise.  Thus, the bond length fluctuates slightly about its average value.
Most of the simulations examined ring polymers, though in some cases linear polymers were also considered.

The polymers are confined to a hard cylindrical tube of diameter $D$. Thus, each monomer interacts
with the wall of the tube with a potential $u_{\rm w}(r) = 0$ for $r<D$ and $u_{\rm w}(r) = \infty$ 
for $r>D$, where $r$ is the distance of the monomer centre from the central axis of the cylinder. Thus, 
$D$ is defined to be the diameter of the cylindrical volume accessible to the centres of the monomers and 
the actual diameter of the cylinder is $D+\sigma$. We consider both infinite- and finite-length tubes. 
In the latter case, each end of the cylinder is capped with a hemisphere whose diameter is equal to 
that of the cylinder. The length, $L$, of the capped tube is defined be that of the cylindrical portion 
of the confinement volume. 

In some simulations, the confining cylinder was also occupied by mobile crowding agents, which were
modeled as hard spheres of diameter $\sigma_{\rm c}$. In this study, we consider crowder sizes 
in the range $0.5\sigma \leq \sigma_{\rm c} \leq \sigma$. In some cases, we consider effectively
infinite cylinder length by employing periodic boundary conditions along the $z$ direction.
The polymer/crowder system is characterized by the crowder packing fraction, 
$\phi_{\rm c} = \pi \sigma_{\rm c}^3/(6V_{\rm cyl})$, where the $V_{\rm cyl}$ is the 
volume of the confining spherocylinder or cylinder, in the case of where periodic boundary conditions
are used.  

The simulations measure the free energy as a function of $\lambda$, the distance between
the polymer centres of mass along the $z$ direction. In addition, the variation of the
overlap length, $L_{\rm ov}$, and the chain extension length, $L_{\rm ext}$, with $\lambda$
are also examined.

\begin{figure}[!ht]
\begin{center}
\includegraphics[width=0.4\textwidth]{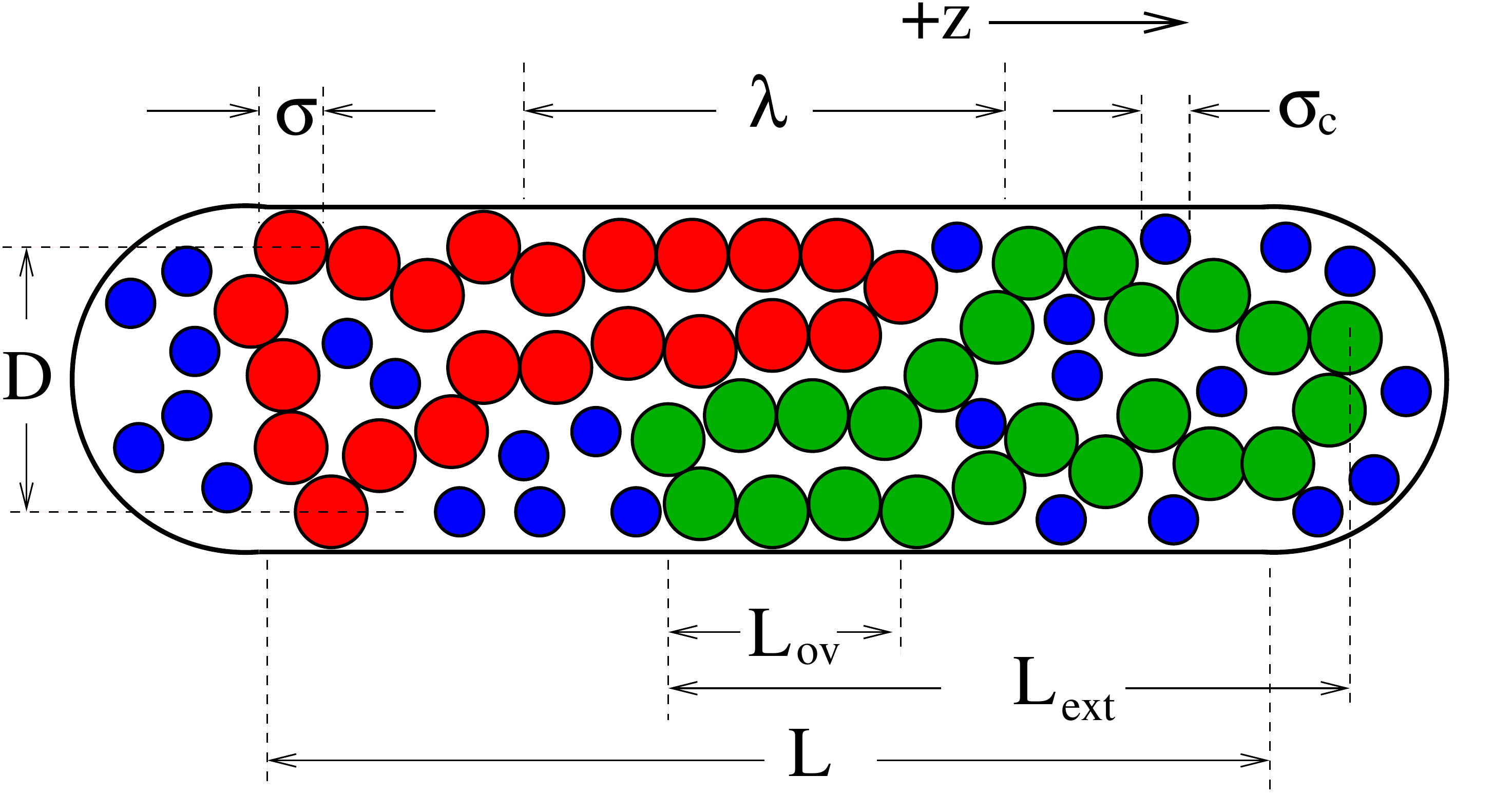}\\
\vspace*{0.2cm}
\includegraphics[width=0.4\textwidth]{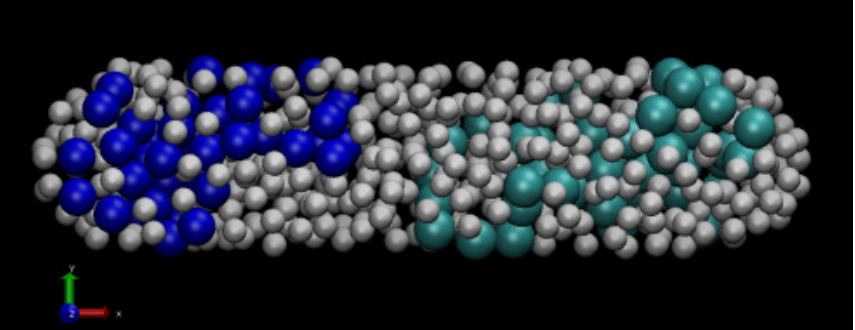}
\end{center}
\caption{
Top: Schematic illustration of the confined two-polymer system showing the definitions 
of the various system parameters described in the text.
Bottom: Snapshot of a system of two confined ring polymers in a cylinder of finite length
in the presence of crowding agents. This image was generated using VMD\cite{HUMP96}
with data taken from a simulation carried out with parameters $N$=40, $L$=14, $D$=4, 
$\sigma_{\rm c}$=0.5, and $\phi_{\rm c}$=0.209.  }
\label{fig:snapshot}
\end{figure}

An illustration of the model system showing the definitions of the various parameters is shown
in Fig.~\ref{fig:snapshot}, along with a snapshot from a simulation.

\section{Methods}
\label{sec:methods}

Monte Carlo simulations employing the Metropolis algorithm and the self-consistent
histogram (SCH) method\cite{frenkel2002understanding} were used to calculate the free energy 
functions for the confined-polymer model system described in Section~\ref{sec:model}. The SCH 
method efficiently calculates the equilibrium probability distribution
${\cal P}(\lambda)$, and thus its corresponding free energy function, 
$F(\lambda) = -k_{\rm B}T\ln {\cal P}(\lambda)$.
We have previously used this procedure to measure free energy functions in our previous
study of polymer segregation,\cite{polson2014polymer} as well as in simulation studies 
of polymer translocation\cite{polson2013simulation,polson2013polymer,polson2014evaluating,%
polson2015polymer} and backfolding of confined polymers.\cite{polson2017free}

To implement the SCH method, we carry out many independent simulations, each of which employs a
unique ``window potential'' of the form:
\begin{eqnarray}
{W_i(\lambda)}=\begin{cases} \infty, \hspace{8mm} \lambda<\lambda_i^{\rm min} \cr 0,
\hspace{1cm} \lambda_i^{\rm min}<\lambda<\lambda_i^{\rm max} \cr \infty, 
\hspace{8mm} \lambda>\lambda_i^{\rm max} \cr
\end{cases}
\label{eq:winPot}
\end{eqnarray}
where $\lambda_i^{\rm min}$ and $\lambda_i^{\rm max}$ are the limits that define the range 
of $\lambda$ for the $i$-th window.  Within each window of $\lambda$, a probability 
distribution $p_i(\lambda)$ is calculated in the simulation. The window potential width,
$\Delta \lambda \equiv \lambda_i^{\rm max} - \lambda_i^{\rm min}$, is chosen to be 
sufficiently small that the variation in $F$ does not exceed a few $k_{\rm B}T$. 
The windows are chosen to overlap with half of the adjacent window, such that 
$\lambda^{\rm max}_{i} = \lambda^{\rm min}_{i+2}$.  The window width was typically 
$\Delta \lambda = 2\sigma$. The SCH algorithm was employed to reconstruct the unbiased 
distribution, ${\cal P}(\lambda)$ from the $p_i(\lambda)$ histograms.  The details of the histogram 
reconstruction algorithm are given in Ref.~\citen{frenkel2002understanding}.  

Polymer configurations were generated by carrying out single-monomer moves using a combination of
translational displacements and crankshaft rotations. In addition, whole-polymer displacements 
of each polymer along the $z$ axis were also carried out to increase the efficiency of
sampling $p_i(\lambda)$.  Each trial move was rejected if it resulted in overlap 
between particles or between a particle and a confinement surface; otherwise it was
accepted.  Initial polymer configurations were generated such that $\lambda$ 
was within the allowed range for a given window potential. Prior to data sampling, the system 
was equilibrated.  As an illustration, for a $N=200$ polymer chain, the system was equilibrated 
for typically $\sim 10^7$ MC cycles, following which a production run of $\sim 10^8$
MC cycles was carried out.  On average, during each MC cycle a displacement or rotation
move for each monomer as well as a whole-polymer displacement along $z$ is attempted once.

In the results presented below, quantities of length are measured in units of $\sigma$ and
energy in units of $k_{\rm B}T$.

\section{Results}
\label{sec:results}

\begin{figure}[!ht]
\begin{center}
\includegraphics[width=0.45\textwidth]{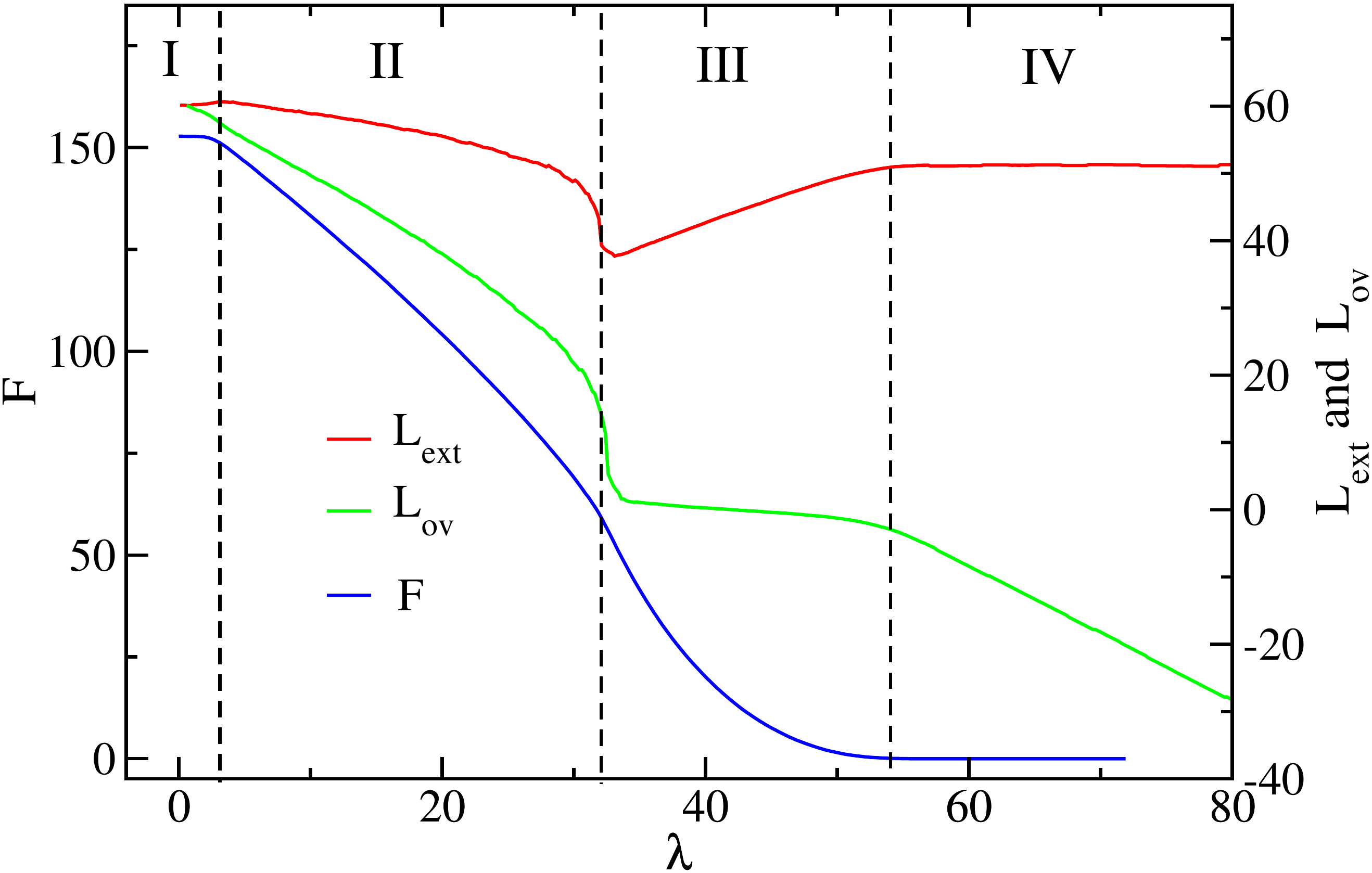}
\end{center}
\caption{Free energy vs $\lambda$ for polymer rings of length $N$=200 in a confining
cylindrical tube of diameter, $D$=4.  The polymer overlap length $L_{\rm ov}$ and 
extension length $L_{\rm ext}$ vs $\lambda$ for the same simulation are overlaid on
the graph. Four different regimes are labeled and are described in the text.}
\label{fig:F.ring.R2.5}
\end{figure}

Consider first the case of two polymers confined to a cylinder of infinite length in the
absence of crowding agents, i.e. $L$=$\infty$ and $\phi_{\rm c}$=0.
Figure~\ref{fig:F.ring.R2.5} shows the variation of $F$ with centre-of-mass separation
distance $\lambda$ for a pair of ring polymers of length $N$=200 in a tube of diameter $D$=4. 
The figure also shows the variation of the polymer overlap length $L_{\rm ov}$ and the extension 
length $L_{\rm ext}$ with $\lambda$ for the same system.

There are four distinct regimes for the functions, which we designate I, II, III and IV,
as labeled in Fig.~\ref{fig:F.ring.R2.5}. 
In regime~IV, the separation $\lambda$ is sufficiently large that the polymers are not in
contact. Consequently, varying $\lambda$ does not change the conformational entropy and 
thus both $F(\lambda)$ and $L_{\rm ext}$ are constant. As $\lambda$ further decreases, the 
polymers make contact, reducing the number of accessible conformations.  Thus, the entropy 
decreases and $F$ rises. As the polymers are initially brought together, the overlap
distance remains $L_{\rm ov}\approx$0 and the extension length $L_{\rm ext}$ decreases.
Thus, in regime~III the polymers are in contact but not overlapping along $z$ and
are compressed. At $\lambda\approx$30, $L_{\rm ext}$ and $L_{\rm ov}$ abruptly increase,
and then continue to increase gradually as $\lambda$ decreases further. This new regime,
i.e. regime~II, is marked by a change in the curvature of $F(\lambda)$. Here, the polymers
overlap along the $z$ axis and the degree of overlap increases with decreasing $\lambda$, 
but at a slower rate than in regime~II and with negative curvature. Finally, at very low 
values of the separation distance, i.e.  $\lambda\leq$4, the system enters regime~I, which
is characterized by constant $F$ and a slight decrease in the average extension length. 
This corresponds to a state where one polymer is nested within the other, as will be clarified 
below.  The trends in $F$, $L_{\rm ov}$ and $L_{\rm ext}$ for the ring polymers are identical
to those observed previously for linear polymers.\cite{polson2014polymer}

\begin{figure}[!ht]
\begin{center}
\includegraphics[width=0.4\textwidth]{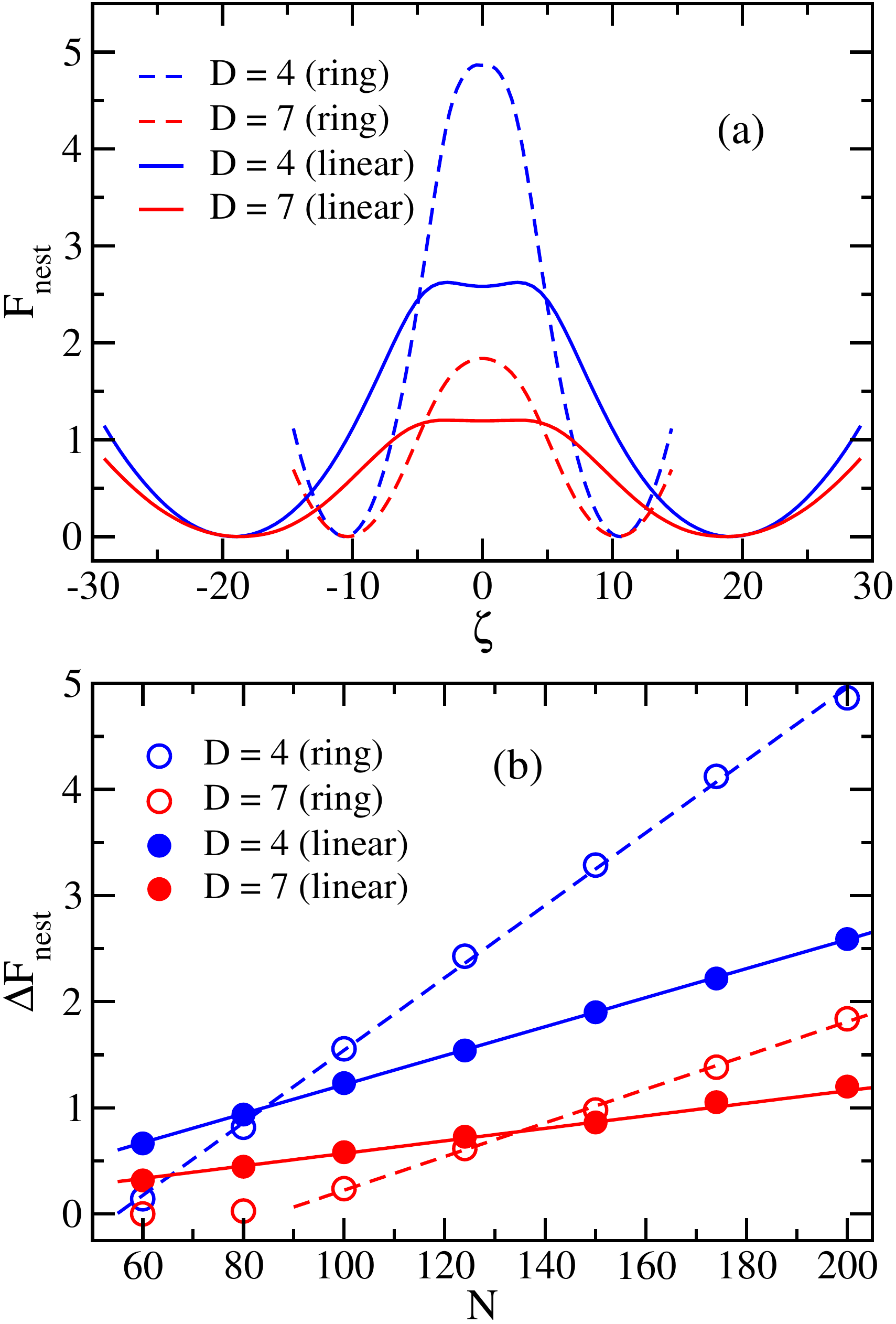}
\end{center}
\caption{(a) Free energy vs polymer extension length difference, $\zeta\equiv 
\Delta L_{\rm ext}$ for $N$=200 polymers with overlapping centres of mass ($\lambda$=0). 
Results are shown for tube diameters of $D$=4 and 7 for linear and ring polymers.
(b) Nesting barrier height $\Delta F_{\rm nest}$ vs chain length $N$ for polymers
with overlapping centres of mass ($\lambda$=0) for $D$=4 and 7 for linear and ring
polymer systems.
}
\label{fig:Fnest}
\end{figure}

Let us first examine the case of regime~I. Here, the centres of mass of the polymers are 
close together and one polymer tends to nest within the other. At the lowest free energy state 
the polymer extension length difference $\zeta\equiv \Delta L_{\rm ext}$ is nonzero.  In 
Fig.~\ref{fig:Fnest}(a), free energy functions $F(\zeta)\equiv -\ln {\cal P}(\zeta)$ are 
plotted for $N$=200 polymers with perfectly overlapping centres of mass, i.e. $\lambda$=0. 
Results are shown for ring and linear polymer systems for $D$=4 and 7. Due to the symmetry 
of the system, $F(-\zeta) = F(\zeta)$, and two equivalent minima are separated 
by a free energy barrier at $\zeta$=0.  The nesting free energy barrier height, 
$\Delta F_{\rm nest}\equiv F_{\rm nest}(0)-F_{\rm nest}(\zeta_{\rm min})$,
is a measure of the degree of preference for the nesting configurations over those
for with equal extension lengths at $\lambda$=0. Generally, $\Delta F_{\rm nest}$
increases linearly with $N$. The rate of increase, $d\Delta F_{\rm nest}/dN$, is greater for 
ring than for linear polymers. In addition, $d\Delta F_{\rm nest}/dN$ increases as the
tube diameter decreases. Finally, the equilibrium chain extension length difference 
$\zeta_{\rm min}$ is only weakly affected by varying $D$, though it is significantly 
greater for linear polymers than for rings.

Minina and Arnold have studied the effects of the nesting free energy barrier on the 
early stages of segregation of initially overlapping polymers.\cite{minina2014induction,%
minina2015entropic} They identify an induction phase in which the system surmounts
the barrier in an activated process in order to break the symmetry of the system
and begin segregation, and they derive expressions that predict the 
form and scaling of $F_{\rm nest}(\zeta)$. The theoretical model employs two
main theoretical tools: (1) the de~Gennes blob model and (2) the approximation that
the confinement free energy of two overlapping chains is equal to the that for
two isolated polymers confined to cylinders of half the cross-sectional area.\cite{jung2012ring}
We will review these concepts in detail and employ them in our analysis of other 
results below. From their analysis, Minina and Arnold predict that 
$\Delta F_{\rm nest}\propto n_{\rm blob} = N D^{-1/\nu}\approx ND^{-1.70}$, where 
$n_{\rm blob}$ is the number of de~Gennes blobs for a single polymer of length $N$
in a tube of diameter $D$, and where $\nu\approx 0.588$ is the Flory exponent. 
The free energy minima are expected to scale 
$\zeta_{\rm min}\propto ND^{1-1/\nu}\approx ND^{-0.70}$. 
In addition, they predict that ring and linear 
polymer barrier heights scale $\Delta F^{\rm (ring)}_{\rm nest} = 2^{1/2\nu} 
\Delta F^{\rm (lin)}_{\rm nest}\approx 1.80 \Delta F^{\rm (lin)}_{\rm nest}$,
and it can further be shown that their analysis predicts 
$\zeta_{\rm min}^{\rm (ring)} = 2\zeta_{\rm min}^{\rm (lin)}$. 
From Fig.~\ref{fig:Fnest} the linear relation between $\Delta F_{\rm nest}$
and $N$ holds, but only when $N$ exceeds approximately $g\approx (D/\sigma)^{1/\nu}$,
the number of monomers in a blob. For example, the $\zeta$-intercept for the data $D=4$ 
for linear polymers is $\zeta\approx 11$, which is equal to $g=4^{1/0.588}$.
The predicted barrier scaling rate $dF_{\rm nest}/dN\propto D^{-1.70}$ suggests
that the ratio of the slopes for $D$=4 and 7 are $(4/7)^{-1.70}=2.6$. This is
somewhat larger than the measured ratio for the slopes of 2.2 for both linear and 
ring polymers. The prediction that $\zeta_{\rm min}\propto N$ is consistent with
the simulation results (data not shown). On the other hand, the scaling
$\zeta_{\rm min} \propto D^{-0.70}$ predicts $\zeta_{\rm min}(D=7)/\zeta_{\rm min}(D=4)
=(7/4)^{-0.7}=0.68$, while the curves in Fig.~\ref{fig:Fnest} clearly show much
smaller shifts in $\zeta_{\rm min}$ with increasing $D$. Finally, the ratio
of the slope $dF_{\rm nest}/dN$ for ring and linear polymers is predicted to be
$2^{1/2\nu}\approx 1.8$, which is noticeably less than the measured value of 2.5
for both $D$=4 and $D$=7.

To summarize, the predicted scaling of $F_{\rm nest}(\zeta)$ is generally accurate
for variation of $N$, but leads to more significant quantitative inconsistencies 
for variations in tube diameter $D$. The prediction for the relation between the
free energy functions for linear and ring polymers suffers to a comparable degree.
The good and poor quality of the predicted scaling with $N$ and $D$, respectively,
was observed previously\cite{minina2014induction,minina2015entropic} and explained
in terms of the limitations validity of the de~Gennes blob model predictions for 
low $g$ and $n_{\rm blob}$ (and therefore low $D$ and $N$).\cite{kim2013elasticity} 
It is likely that this at least partially explains the discrepancies seen here.
For example, in the case of linear polymers for $D$=4 and 7, $\zeta_{\rm min}\approx 10$,
i.e. the length of the non-overlap portions of each polymer is only about
$\zeta_{\rm min}/2$=5. This is comparable to the tube diameter, which implies that there
is only about one blob in each non-overlap region. Given this fact, finite-size
effects are hardly surprising.For much larger systems, the number of such blobs will 
be larger, and consequently the resulting finite-size effects are expected to be smaller.

It is noteworthy that regime~I is characterized by a free energy function 
$F_{\rm nest}(\zeta)$ with a barrier that increases with system size, and is
also characterized by that part of $F(\lambda)$ which is constant with respect
to $\lambda$, independent of system size. This suggests two possible dynamical
pathways during the early stages of segregation starting from $\lambda$=0. One 
route is that suggested by Minina and Arnold\cite{minina2014induction,minina2015entropic} 
in which a fluctuation causes one half of the nested polymer to overcome the barrier
in $F(\zeta)$, break the symmetry, and enter regime~II where the gradient in $F(\lambda)$
then drives segregation. In the other scenario, the polymer remains in a nested state,
but its centre of mass diffuses relative to the other until one end of the polymer
reaches that of the other polymer, after which the system jumps into regime~II. It is
by no means obvious which mechanism (or some combination) is relevant for any given
set of system parameters. The preferred route will likely depend on the barrier height 
$\Delta F_{\rm nest}$, as well as on the diffusion coefficients associated with each
process, and all these quantities depend on $N$, $D$ and polymer topology. 
%
%

\begin{figure}[!ht]
\begin{center}
\includegraphics[width=0.43\textwidth]{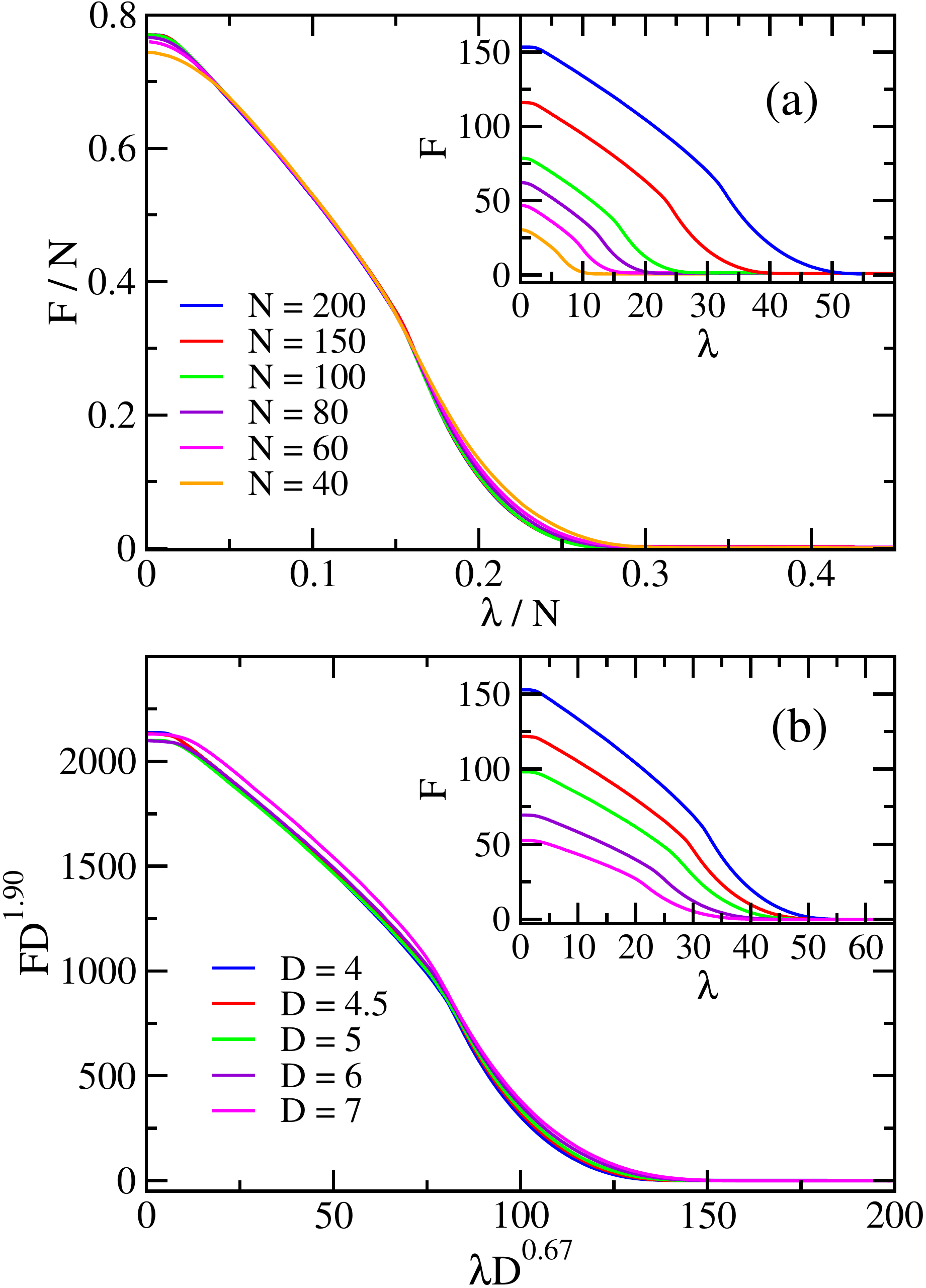}
\end{center}
\caption{(a) Scaled free energy functions for systems of polymer rings confined to an
infinitely long cylinder of diameter $D$=4. Results for various values of the polymer
length $N$ are shown.  The inset shows shows unscaled functions using the same data.
(b) Scaled free energy functions rings of length $N$=200 in a confining cylindrical tube 
for several values of the tube diameter, $D$. The inset unscaled free energy functions 
using the same data. }
\label{fig:F.ring.NDscale}
\end{figure}

Next, we consider the scaling behaviour of the system for the other regimes.
Figure~\ref{fig:F.ring.NDscale} illustrates the effects of varying $N$ and the tube diameter
$D$ on the free energy functions. The insets of the Fig.~\ref{fig:F.ring.NDscale}(a)
and (b) show the measured free energy functions for various $N$ and $D$, respectively.
The trends are straightforward. The free energy barrier height, $\Delta F\equiv F(0)-F(\infty)$,
decreases monotonically with increasing $N$ and $D$. In addition, increasing $N$ and $D$
reduces the range of $\lambda$ over which the polymers are in contact.  This
latter feature is simply a result of the reduction of the extension length $L_{\rm ext}$
of the polymer along $z$ as $N$ and $D$ each increase (data not shown); that is, as
$L_{\rm ext}$ decreases, the polymers must be closer together before intermolecular contact
has the effect of reducing the conformational entropy and increasing $F$.  The main parts of 
Fig.~\ref{fig:F.ring.NDscale}(a) and (b) show the quantitative scaling results for $F(\lambda)$.
Figure~\ref{fig:F.ring.NDscale}(a) shows that curves for $F/N$ vs $\lambda/N$ at fixed $D$ 
tend to collapse onto a universal curve, while Fig.~\ref{fig:F.ring.NDscale}(b) shows that
plotting $FD^{1.90}$ vs $\lambda D^{0.67}$ also produces curves that collapse onto a single
curve. Small deviations are noted for low $N$ and large $D$, but otherwise, the data collapse
is very good. Together, these results suggest that the free energy functions are of the form: 
\begin{eqnarray}
F(\lambda; N,D) = N D^{-\alpha} f(\lambda/ ND^{-\beta})
\label{eq:FlND}
\end{eqnarray}
where $f(x)$ is a universal function that satisfies $f(\infty)=0$ and where the exponents 
have values $\alpha$=1.90 and $\beta$=0.67.

\begin{figure}[!ht]
\begin{center}
\includegraphics[width=0.43\textwidth]{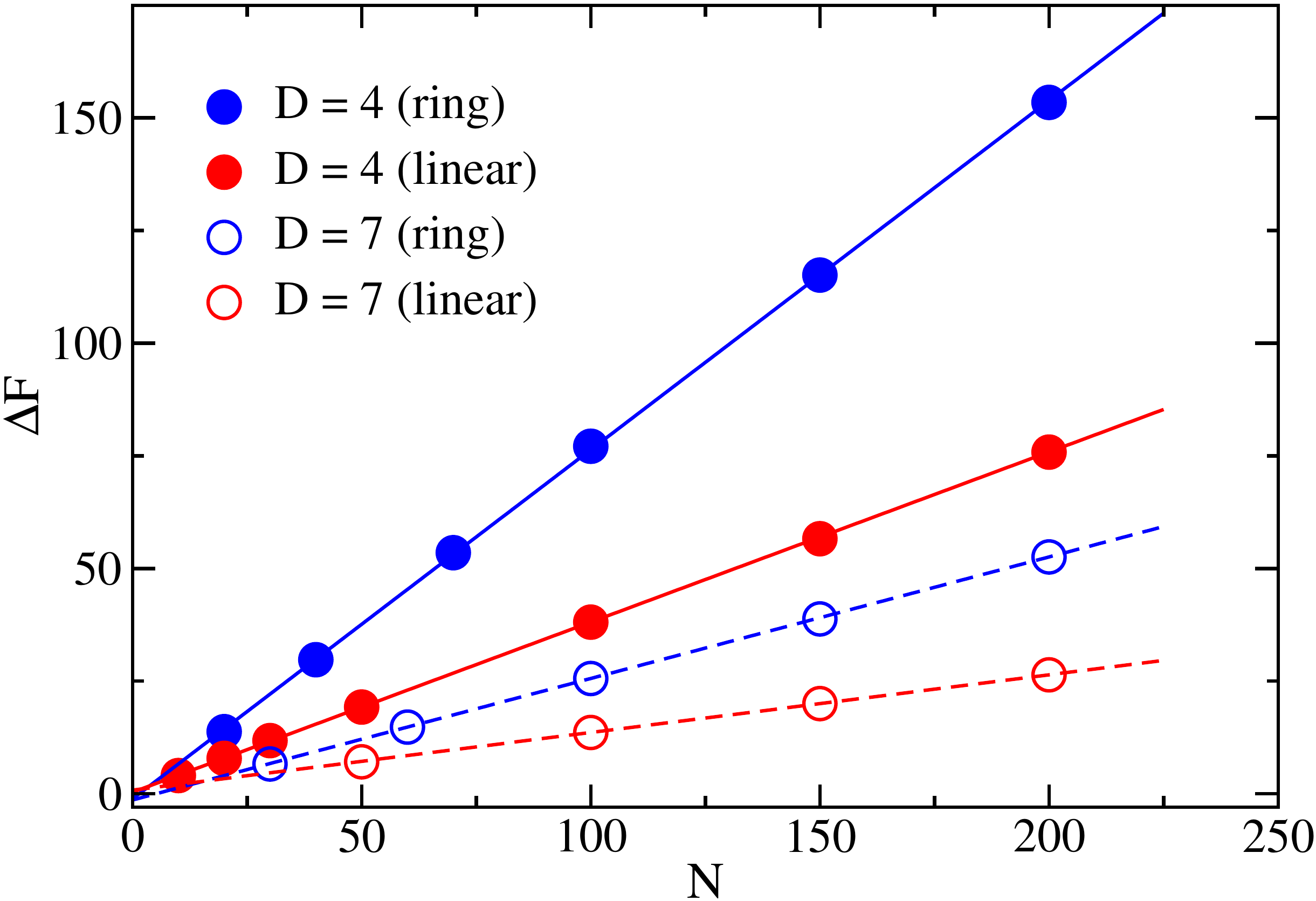}
\end{center}
\caption{Comparison of the variation of the free energy barrier height $\Delta F$ with 
polymer length $N$ for ring polymers and linear polymers, each for $D$=4 and $D$=7.
}
\label{fig:delF.N.compare}
\end{figure}

Let us now compare the scaling properties of $F(\lambda)$ for ring polymers and linear polymers.
Figure~\ref{fig:delF.N.compare} compares the variation in the free energy barrier, 
$\Delta F \equiv F(0)-F(\infty)$, with polymer length $N$ for ring polymers and linear polymers.
Results for tube diameters of $D$=4 and $D$=7 are shown. In each case, $\Delta F$ is proportional
to $N$. Note that this is consistent with the more general result of Eq.~(\ref{eq:FlND}), since
$\Delta F\sim ND^{-\alpha} f(0)$. In addition, this scaling was noted earlier for linear
polymers in Ref.~\citen{polson2014polymer}.  For $D$=4, fits to the data yield values of the 
barrier height per monomer of $\Delta F/N$=0.799 for ring polymers and 0.378 for linear polymers. 
For $D$=7, we find $\Delta F/N$=0.263 for ring polymers and 0.132 for linear polymers.

\begin{figure}[!ht]
\begin{center}
\includegraphics[width=0.4\textwidth]{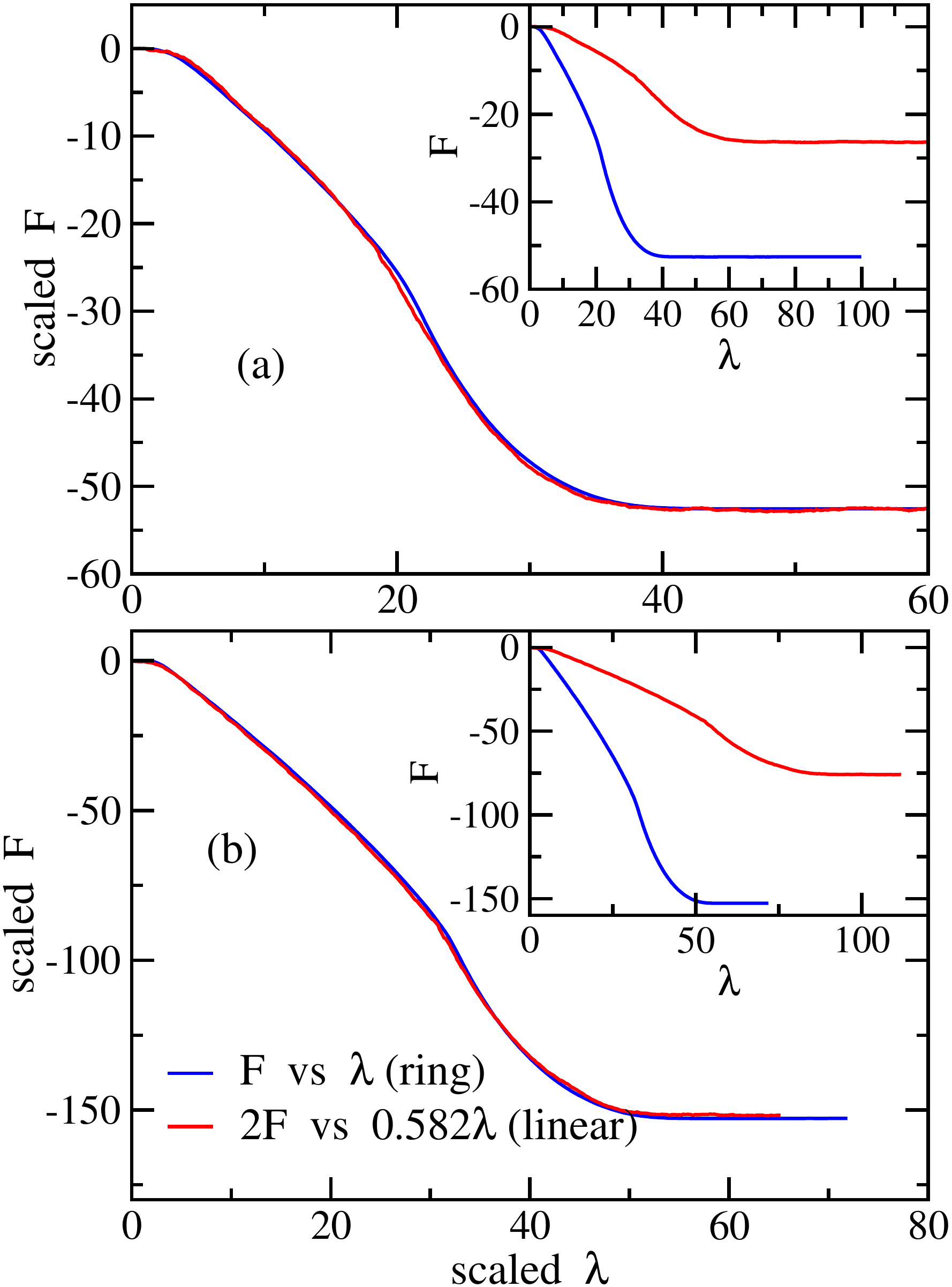}
\end{center}
\caption{(a) The blue curve shows the overlap free energy vs $\lambda$ for polymer 
rings of length $N=200$ in a confining cylindrical tube of diameter $D=7$. 
The overlaid red curve shows the scaled overlap free energy function of two cylindrically
confined $N=200$ linear polymers. The latter curve has been scaled along $F$ and $\lambda$,
as indicated in the legend. The inset shows unscaled free energy functions for the
ring and linear polymer systems. (b) As in (a), except for a confining cylinder diameter
of $D=4$.  }
\label{fig:F.ring.linear.N200}
\end{figure}

Figure~\ref{fig:F.ring.linear.N200} compares free energy functions for ring and linear
polymers of length $N$=200 for $D$=4 in (a) and for $D$=7 in (b). The insets show the
unscaled functions for each case. While the qualitative features are the same for both 
polymer topologies, as noted earlier, the quantitative results differ appreciably.
Generally, for the same $N$ and $D$, ring polymers have higher free energy barriers.
In addition, the range of $\lambda$ for regime~IV extends to lower $\lambda$ for
rings compared to linear polymers. The explanation for the latter trends is straightforward.
The mean extension length $L_{\rm ext}$ is shorter for rings than for linear polymers,
and consequently the polymers must get closer (i.e. lower $\lambda$) before the
polymers make contact and $F$ rises.

A prediction for the scaling of $F(\lambda)$ with $N$ and $D$ for both rings and linear chains that 
is consistent with the form of Eq.~(\ref{eq:FlND}) can be obtained using simple scaling arguments.
Let us consider first the case of a system of linear polymers. In regime~II, each polymer 
has $N_{\rm ov}$ monomers in the overlap region and $N-N_{\rm ov}$ monomers in the non-overlap 
region. Employing the de~Gennes blob model, the length of the non-overlap region is 
$L_{\rm ov}-L_{\rm ext} = c_1(N-N_{\rm ov}) D^{1-1/\nu}$, where $c_1$ is a non-universal constant. 
To determine $L_{\rm ov}$, we employ an approach used by Ra{\v{c}}ko and Cifra\cite{racko2013segregation} 
building on an idea introduced by Jung {\it et al.}\cite{jung2012ring}
When two polymers overlap along the tube, each overlapping portion effectively occupies a tube of
half the cross-sectional area of the real tube. Thus, the effective diameter is $D/\sqrt{2}$.
Consequently, it follows that $L_{\rm ov} = 2^{-1/2+1/2\nu} c_1 N_{\rm ov} D^{1-1/\nu}$. From these
relations, it is readily shown that the centre-of-mass separation distance satisfies
$\lambda = ND^{1-1/\nu} (1-N_{\rm ov}/N)\left((2^{-1/2+1/2\nu}-1)(N_{\rm ov}/N) + 1\right)$.
Inverting the equation it follows that: $N_{\rm ov}/N = u\left(\lambda/ND^{1-1/\nu}\right)$
where $u(x)$ is a function obtained from solving the quadratic equation for $N_{\rm ov}/N$.
The overlap free energy of the two chains is determined by assigning $k_{\rm B}T$ to each blob
in the overlap region for both polymers. Again using an effective diameter of $D/\sqrt{2}$ for 
monomers in this region, it follows that $F^{\rm (lin)} = 2^{1+1/2\nu} N_{\rm ov} D^{-1/\nu}$. 
Using the previous result for $N_{\rm ov}/N$, we find
\begin{eqnarray}
F^{\rm (lin)}(\lambda) = 2^{1+1/2\nu} N D^{-1/\nu} u(\lambda/ND^{1-1/\nu}).
\label{eq:FlinII}
\end{eqnarray}

Next consider regime~III, where the polymers are compressed and in contact, but not overlapping. 
Employing the renormalized Flory theory of Jun {\it et al.},\cite{jun2008compression} the free 
energy for a single linear polymer of length $L_{\rm ext}$ is 
\begin{eqnarray}
F=AL_{\rm ext}^2/(N/g)D^2 + BD(N/g)^2/L_{\rm ext},
\label{eq:Frenorm}
\end{eqnarray}
where $A$ and $B$ are constants of order unity and $g\sim D^{1/\nu}$ is the number of
monomers in a compression blob of diameter $D$. Approximating $L_{\rm ext}\approx\lambda$
(i.e. assuming uniform compression) and noting the equal contributions from the two chains,
it is easily shown that 
\begin{eqnarray}
F^{\rm (lin)}(\lambda) = N D^{-1/\nu} w(\lambda/ND^{1-1/\nu}),
\label{eq:FlinIII}
\end{eqnarray}
where $w(x) = 2(Ax^2+B/x)$. Comparing with Eq.~(\ref{eq:FlND}), 
The transition between regimes~II and III will occur at a separation $\lambda^*$ defined by
$2^{1+1/2\nu} u(\lambda^*/ND^{1-1/\nu})=w(\lambda^*/ND^{1-1/\nu})$. The value of $\lambda^*$ 
is determined by non-universal constants such as $A$, $B$ and $c_1$ that cannot be 
estimated by a scaling analysis.  Note from Eqs.~(\ref{eq:FlinII}) and (\ref{eq:FlinIII}) 
that the scaling of $F(\lambda)$ with $N$ and $D$ are identical for regions II and III.  
In addition, the magnitude of the reduction in $F$ in regime~I due to
nesting effects was also predicted to scale in the same manner, as noted earlier.
Thus, the form of $F(\lambda)$ is consistent with that of Eq.~(\ref{eq:FlND}) in all
three regimes with predicted exponents of $\alpha=1/\nu\approx 1.70$ and 
$\beta=1/\nu-1\approx 0.70$.  These are comparable to, but somewhat less than the 
measured values of $\alpha=1.90$ and $\beta=0.67$.

The predicted scaling behaviour of $F(\lambda)$ is consistent with that derived by
 Minina and Arnold,\cite{minina2014induction,minina2015entropic} who used the same 
approach.  However, in that work regime~III is not accounted for and regime~II
is assumed to persist with increasing $\lambda$ until the polymers are no longer in contact.

A prediction for the scaling of $F(\lambda)$ with $N$ and $D$ for ring polymers follows
a very similar approach as that for linear polymers. As suggested by Jung
{\it et al.},\cite{jung2012ring} a single ring polymer of length $N$ can be modeled as 
two completely overlapping chains of length $N/2$ in a tube of effective
diameter $D/\sqrt{2}$. Thus, for two overlapping ring polymers in regime~II, the
non-overlapping portion of either polymer comprising $N-N_{\rm ov}$ monomers each 
has a length $L_{\rm ext}-L_{\rm ov} = 2^{-3/2+1/2\nu} c_1 (N-N_{\rm ov}) D^{1-1/\nu}$,
where $c_1$ is the same non-universal constant used earlier.
In the overlapping region, there are effectively four subchains occupying effective
tubes of 1/4 the cross-sectional area of the real confining tube; thus, the effective
diameter for each is $D/2$. Using de~Gennes blob scaling for each, the overlap region
has a predicted span of $L_{\rm ov} = 2^{-2+1/\nu} N_{\rm ov} D^{1-1/\nu}$. From these
expressions for the two lengths, the centre-of-mass separation of two rings is easily 
calculated to be
\begin{eqnarray*}
\frac{\lambda}{2^{-3/2+1/2\nu} ND^{1-1/\nu}} = 
\left(1-\frac{N_{\rm ov}}{N}\right)\left((2^{-1/2+1/2\nu}-1)\left(\frac{N_{\rm ov}}{N}\right) 
+ 1\right).
\end{eqnarray*}
Inverting this equation yields
\begin{eqnarray}
\frac{N_{\rm ov}}{N} = u\left(\frac{\lambda}{2^{-3/2+1/2\nu} ND^{1-1/\nu}}\right),
\label{eq:NovNring}
\end{eqnarray}
where $u$ is the same function appearing in Eq.~(\ref{eq:FlinII}). Now, assigning $k_{\rm B}T$
to each blob in the four subchains of the overlapping region yields an overlap free energy of
$F^{\rm (ring)}(\lambda) = ND^{-1/\nu} 2^{1+1/\nu} (N_{\rm ov}/N)$. Using Eq.~(\ref{eq:NovNring}), 
this gives
\begin{eqnarray}
F^{\rm (ring)}(\lambda) = 2^{1+1/\nu} ND^{-1/\nu} u\left(\frac{\lambda}{2^{-3/2+1/2\nu} 
ND^{1-1/\nu}}\right).
\label{eq:FringII}
\end{eqnarray}

Now consider ring polymers in regime~III. Here, each compressed (but non-overlapping) chain
of length $N$ can be modeled as two independent subchains of length $N/2$ in an effective
tube of diameter $D/\sqrt{2}$. Using the prediction from the renormalized Flory theory
of Jun {\it et al.}\cite{jun2008compression} in Eq.~(\ref{eq:FlND}) with 
$L_{\rm ext}\approx\lambda$ and substituting $D\rightarrow D/\sqrt{2}$ and $N\rightarrow N/2$, 
it can be shown that
\begin{eqnarray}
F^{\rm (ring)}(\lambda) = 2^{1/2\nu} 
ND^{-1/\nu} w\left(\frac{\lambda}{2^{-3/2+1/2\nu} ND^{1-1/\nu}}\right).
\label{eq:FringIII}
\end{eqnarray}
where $w(x)\equiv 2(Ax^2+B/x)$. As before, the separation distance $\lambda^*$ dividing
regimes~II and III is determined by the relation 
$2^{1+1/2\nu} u\left(\lambda^*/2^{-3/2+1/2\nu} ND^{1-1/\nu}\right) 
= w\left(\lambda^*/2^{-3/2+1/2\nu} ND^{1-1/\nu}\right)$.
Note that Eqs.~(\ref{eq:FringII}) and (\ref{eq:FringIII}) predict the same scaling of 
$F(\lambda)$ with $N$ and $D$. Thus, as was the case for linear polymers,
$F(\lambda)$ is predicted to scale in the same manner as Eq.~(\ref{eq:FlND}) with the
correct scaling for $N$ and predicted exponents for $D$ of $\alpha=1/\nu\approx 1.70$ and 
$\beta=1/\nu-1\approx 0.70$. Again, these are comparable to, but slightly different from the 
observed values of $\alpha=1.90$ and $\beta=0.67$. In addition, comparing the results for 
linear and ring polymers in Eqs.~(\ref{eq:FlinII}), (\ref{eq:FlinIII}), (\ref{eq:FringII}), 
and (\ref{eq:FringIII}), we see that
\begin{eqnarray}
F^{\rm (ring)}(\lambda;N,D) = 2^{1/2\nu} F^{\rm (lin)}(\lambda/2^{-3/2+1/2\nu};N,D).
\label{eq:Fringlin}
\end{eqnarray}
Thus, for a given $N$ and $D$, the function $F^{\rm (ring)}(\lambda)$ is related 
to $F^{\rm (lin)}(\lambda)$ by a scaling of $2^{1/2\nu}\approx 1.80$ along $F$ and
a scaling of $2^{-3/2+1/2\nu}\approx 0.637$ along $\lambda$. These values are comparable 
to, but slightly different from the factors of $2.0$ and $0.582$, respectively, that
were used required to obtain collapse of the functions for ring and linear polymers
in Fig.~\ref{fig:F.ring.linear.N200}.

As noted above, the predicted scaling of $F(\lambda)$ with $N$ for both ring and linear 
polymers is quantitatively consistent with the predictions. However, the scaling 
with respect to $D$, as well as the predicted relationship between the ring and linear 
polymer systems is somewhat poorer. This discrepancy is expected to arise from the 
inadequacy of either one or both of two approximations.  First, the measured scaling 
exponents of polymers in simulations typically show significant deviations 
from the predictions from the de~Gennes blob model as a result of finite-size effects. 
Recall that each blob, which contributes approximately $k_{\rm B}T$ to the
confinement free energy, consists of $g\sim D^{1/\nu}$ monomers each. Thus, there are 
$n_{\rm blob}=N/g$ blobs per polymer.  In order for these predictions to be accurate, it is 
required that $g\gg 1$ and $n_{\rm blob}\gg 1$. Simulations have revealed that the first
condition requires tube diameters $D\geq$10, while the addition of the second condition 
requires polymer lengths in the range $N\geq 10^3$ for the type of model employed 
here.\cite{kim2013elasticity} Clearly, neither of these conditions is satisfied in the 
simulations carried out in this study. Consequently, finite-size effects are expected.

\begin{figure}[!ht]
\begin{center}
\includegraphics[width=0.4\textwidth]{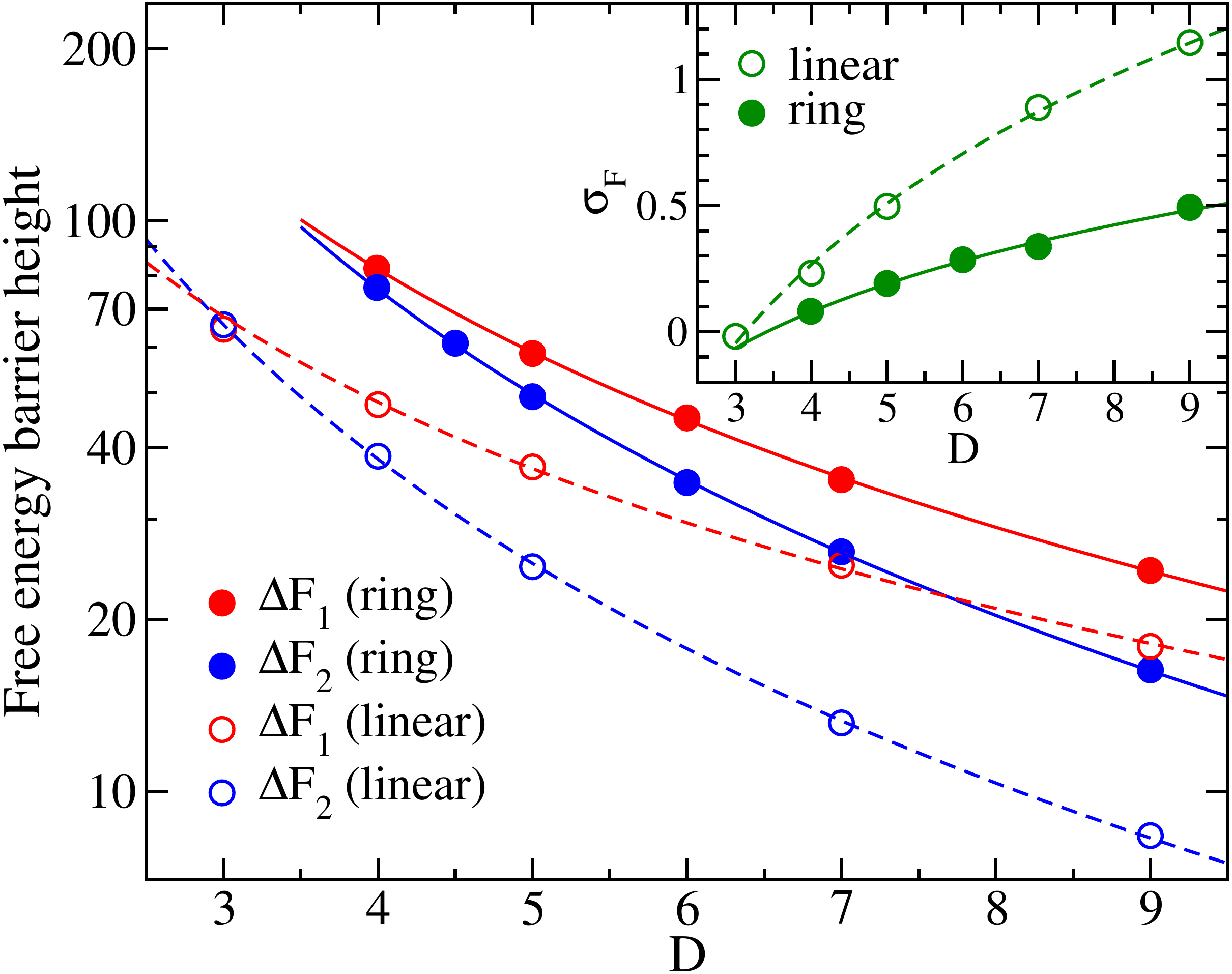}
\end{center}
\caption{
Free energy barrier height per chain vs tube diameter $D$ for polymers of length $N$=200.
The graph compares the overlap free energy barrier height per chain for two polymer chains,
$\Delta F_2\equiv \Delta F/2$ (blue symbols), and the confinement free energy difference 
for a single polymer between tubes of diameter $D$ and $D/\sqrt{2}$, $\Delta F_1$ 
(red symbols). Results are
shown for both ring polymers (closed symbols) and linear polymers (open symbols).
The curves overlaid on the data are fits to a power law, $\Delta F_{\rm c} \sim D^{-\alpha}$.
The inset shows the variation of the relative difference between the two barrier heights,
$\sigma_{\rm F}\equiv (\Delta F_1-\Delta F_2)/\Delta F_2$, vs tube diameter $D$.  }
\label{fig:delF.single.double.N200}
\end{figure}

A second possible problem concerns the hypothesis that two overlapping chains have the
same confinement free energy as two chains each in separate tubes of half the cross-sectional
area as the real confining tube. To our knowledge, the accuracy of this approximation
has not been tested previously. To do so in this study, we have carried out 
additional simulations of a single polymer that transitions between a tube of diameter 
$D$ and another of diameter $D/\sqrt{2}$. The two tube sections were connected by a 
tapered section of finite length. The free energy was measured as a function of the 
centre of mass along the composite tube using the same SCF Monte Carlo method used for
the other calculations. Figure~\ref{fig:delF.single.double.N200} compares the free energy 
difference of the polymer between the two tube sections, $\Delta F_{1}$, and the overlap 
free energy barrier per polymer, i.e. $\Delta F_{2}=\Delta F/2$, for the two-polymer 
system. The free energy differences are each plotted as a function of $D$ for both ring 
and linear polymer systems.  The results show that the approximation overestimates the 
free energy barrier height and that the magnitude of this difference grows with increasing 
$D$. The inset of the figure shows the relative difference,
$\sigma_{\rm F}\equiv (\Delta F_2 - \Delta F_1)/\Delta F_1$,
vs $D$. These results demonstrate that both the absolute and relative difference increases with $D$.
At large $D$, the difference is appreciable, reaching $\sigma_{\rm F}\approx 1$ (i.e. a factor
of two difference) for $D$=9 in the case of linear polymers. At the same $D$, the discrepancy
is not as large for ring polymers, though it does grow to an appreciable $\sigma_{\rm F}\approx 0.5$
for $D$=9. The cause of the overestimation of the confinement free energy in this approximation is
straightforward. When two polymers completely overlap (i.e. when $\lambda$=0) the presence of
the other polymer will reduce the number of accessible conformations. However, polymer
conformations will clearly be accessible in the 2-chain system that will not be accessible
for the 1-chain system with a reduced tube diameter. Essentially this is due to the presence
of a hard confining surface of the latter system in contrast to the lateral interpenetration of
the two overlapping chains in the former. As $D$ decreases, there will likely be less such
interpenetration and the 2-chain system behaves more like two isolated 1-chain systems. Consequently,
$\sigma_{\rm F}$ decreases with decreasing $D$. Likewise, at any given $D$, ring polymers
have a higher degree of crowding than do linear polymers. Thus, there is less interpenetration
of the adjacent chains, and $\sigma_{\rm F}$ is lower.

It is difficult to determine the degree to which the inadequacies of these two approximations 
contribute to the discrepancies between the predicted and observed scaling of $F(\lambda)$. 
It is possible that one dominates, or even that there is a cancellation of opposing effects. 
In any case, it is worth considering the expected validity of the predictions for systems of 
very large $N$ and $D$.  In this regime, finite-size effects associated with the de~Gennes blob 
model are expected to vanish, and the predicted scaling exponents for confined chains recovered.
However, it is unlikely that the inaccuracy of the approximation used to calculate the
overlap free energy for overlapping chains will also vanish. Consequently, it is doubtful
whether the scaling of Eq.~(\ref{eq:FlND}) with exponents of $\alpha=1/\nu$ and $\beta=1/\nu-1$ 
will emerge even in the limit of large $N$ and $D$, where $g\sim D^{1/\nu}\gg 1$ and 
$n_{\rm blob}=N/g\gg 1$ are both satisfied. Confirmation of this requires simulations using system 
sizes that are currently not computationally feasible.

\begin{figure}[!ht]
\begin{center}
\includegraphics[width=0.4\textwidth]{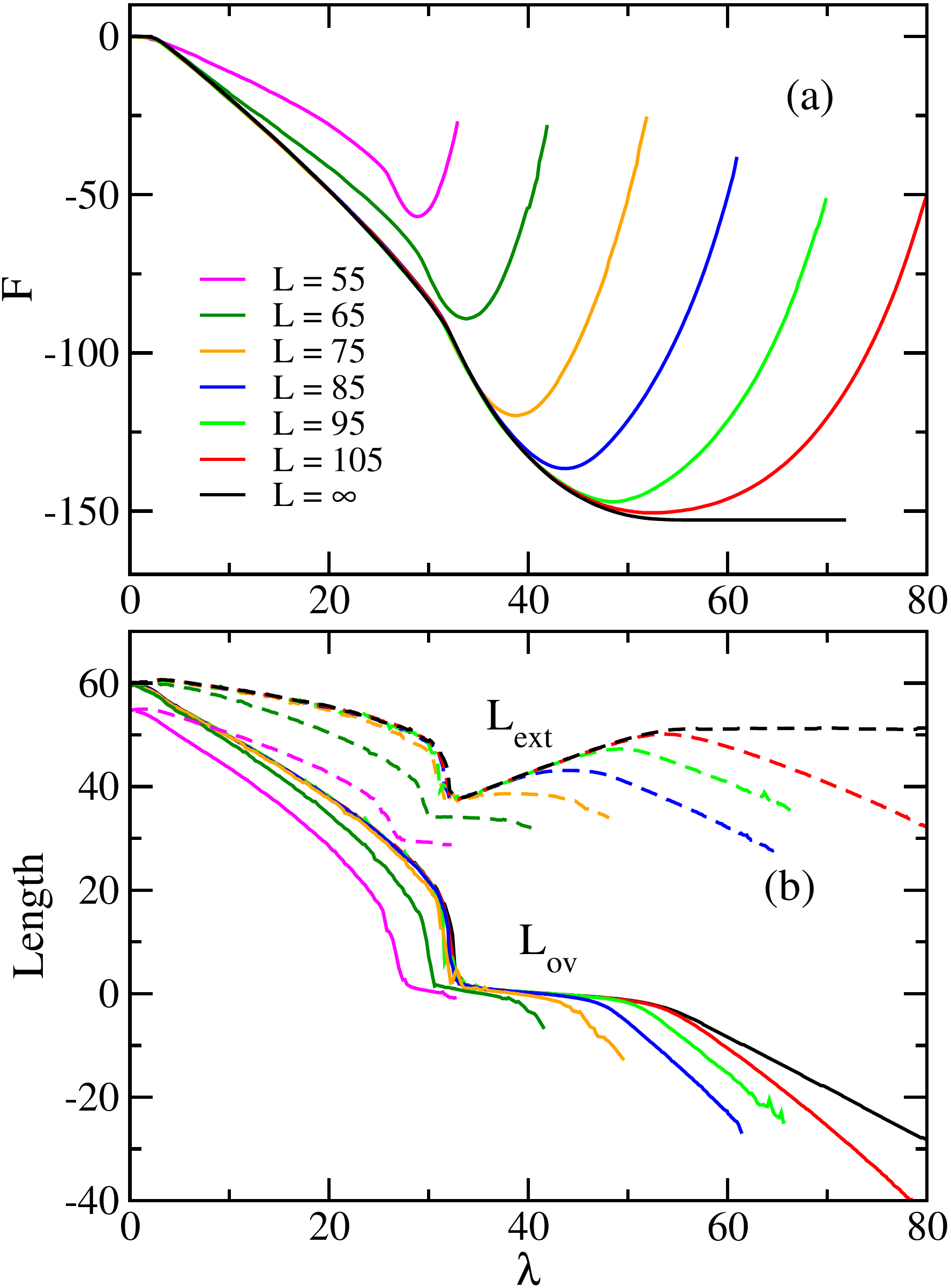}
\end{center}
\caption{(a) Free energy vs $\lambda$ for polymer rings of length $N=200$ in a 
confining cylindrical tube of diameter $D=4$. Results for several values of the 
cylinder length, $L$, are shown. (b) Overlap length $L_{\rm ov}$ (solid curves)
and extension length $L_{\rm ext}$ (dashed curves) vs $\lambda$ for the same systems 
as in (a).  }
\label{fig:F.lover.ring.cap.N200.R2.5}
\end{figure}

Let us now consider polymer systems confined to cylinders of finite length with
hemispheric end caps. Figure~\ref{fig:F.lover.ring.cap.N200.R2.5}(a) shows free energy
functions for a system of ring polymers of length $N$=200 confined to a cylinder of 
diameter $D$=4 for several different values of cylinder length $L$. A curve for $L=\infty$ 
is also shown for comparison. The trends are qualitatively consistent with previous 
calculations for linear polymers.\cite{polson2014polymer} The main new feature for finite 
$L$ that is not present for $L=\infty$ is a steep rise in $F$ at larger values of $\lambda$.
This is due to the fact that the polymers press up against the end-caps of the
confining cylinder at large separation. Consequently, there is a reduction in the
number of accessible conformations, leading to a decreasing entropy. The resulting
increase in $F$ leads to a minimum in $F$, which marks the location of the most
probable separation distance. The longitudinal compression of the polymers at
large $\lambda$ is evident by the decrease in $L_{\rm ext}$ with increasing $\lambda$
in this regime, which is shown in Fig.~\ref{fig:F.lover.ring.cap.N200.R2.5}(b).

\begin{figure}[!ht]
\begin{center}
\includegraphics[width=0.4\textwidth]{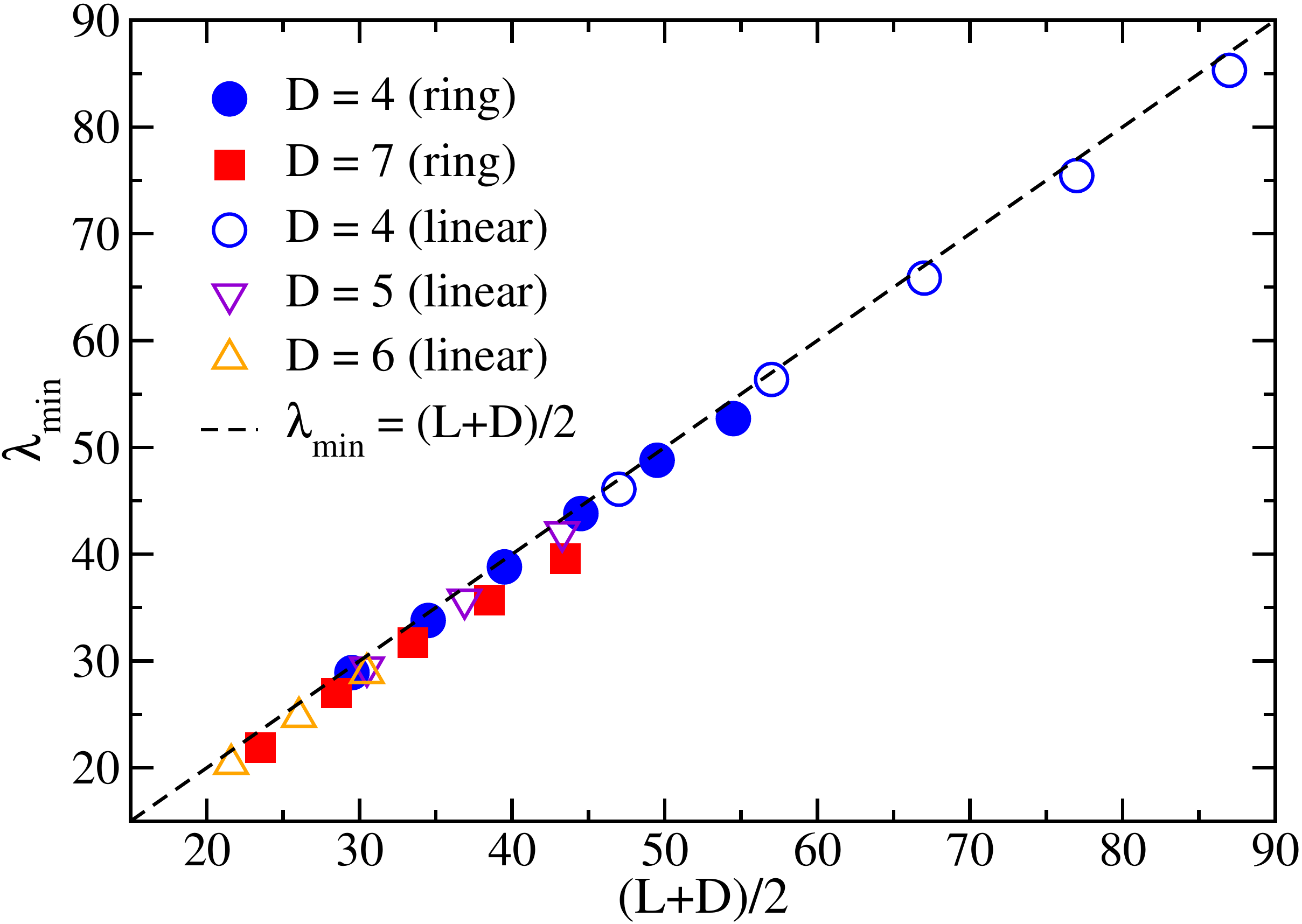}
\end{center}
\caption{Free energy minimum centre-of-mass separation $\lambda_{\rm min}$ vs $(L+D)/2$ 
for polymers confined to cylinders of finite length. Results are shown for ring and linear 
polymers of length $N$=200 for various values of the confinement diameter $D$. } 
\label{fig:lambda_min}
\end{figure}

At sufficiently low $\lambda$, the free energy functions overlap with that of $L$=$\infty$.
The location where a finite-$L$ curve peels off the $L$=$\infty$ curve corresponds
to the separation $\lambda$ where the polymer first makes appreciable contact with the
end-walls. As expected, these $\lambda$ values decrease with decreasing $L$.
The same trend is evident for $L_{\rm ext}$ and $L_{\rm ov}$ in 
Fig.~\ref{fig:F.lover.ring.cap.N200.R2.5}(b). The notable exception is the curve for
$L$=55. In this case, we note that for completely overlapping polymers, i.e. $\lambda$=0,
the measured extension length $L_{\rm ext}\approx 55$ is lower than the value for $L$=$\infty$ 
of $L_{\rm ext}\approx 60$. Thus, polymers in such a sufficiently short cylinder already
feel the effects of longitudinal confinement in a completely overlapping configuration.
In all cases, the free energy minimum lies in a regime where the polymers are
in regime~II. Thus, the polymers are segregated but compressed by contact with
the second polymer on one side and by the hemispheric end-cap on the other.
In addition, the centre-of-mass separation $\lambda$ at the minimum is expected to be 
approximately half the total length of the cylinder, i.e. 
$\lambda_{\rm min}\approx (L+D)/2$. Figure~\ref{fig:lambda_min} shows results
for $\lambda_{\rm min}$ vs $(L+D)/2$ for both ring and linear polymer systems
for $N$=200 polymers with various values of $D$. The behaviour is in good quantitative 
agreement with this prediction, though there is a small but consistent overestimate. 
This is likely due to the combined effects of the somewhat stronger lateral confinement at 
the hemispheric end-cap as well as a small degree chain overlap at the interface.

\begin{figure}[!ht]
\begin{center}
\includegraphics[width=0.4\textwidth]{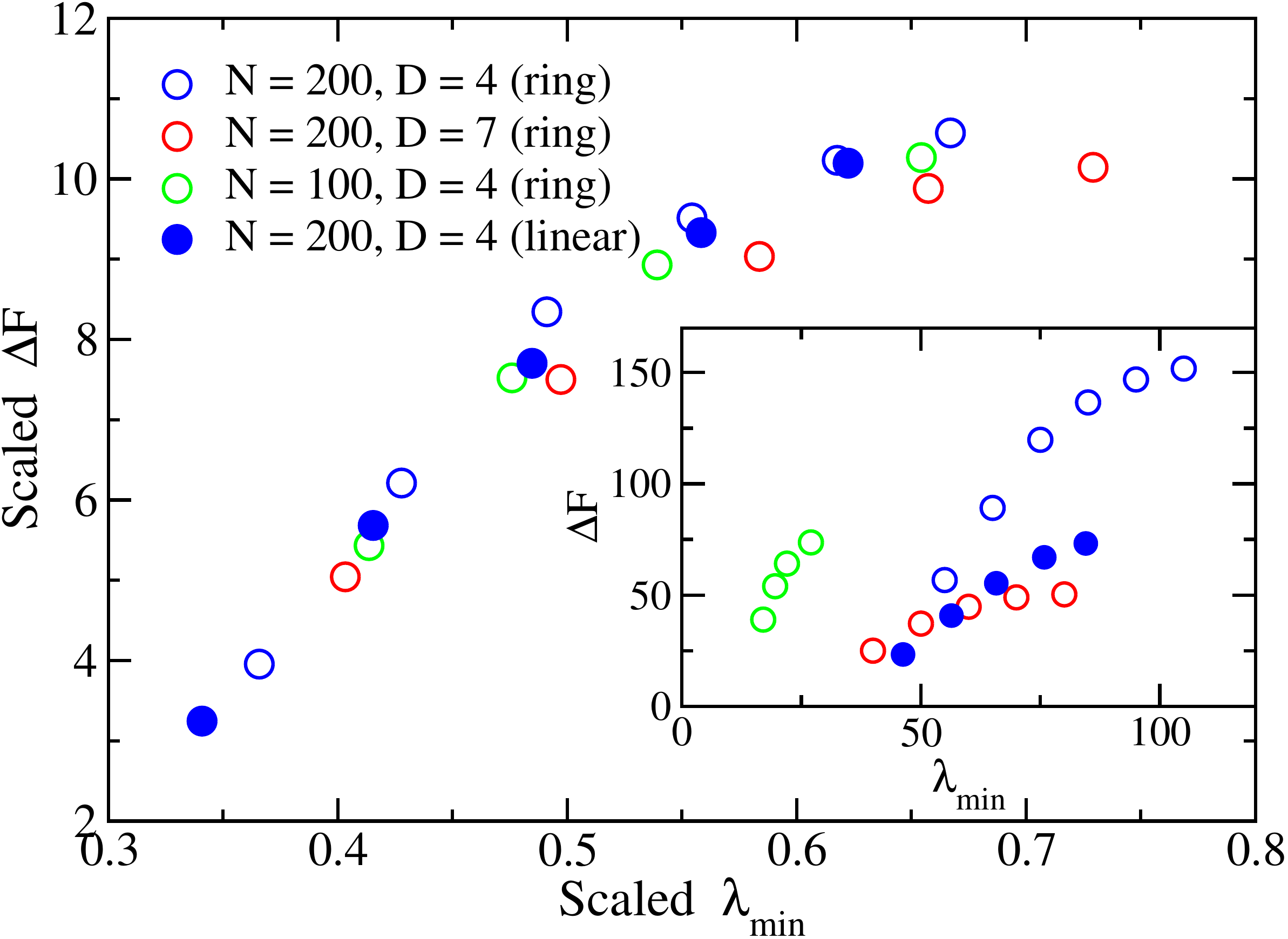}
\end{center}
\caption{Scaled free energy barrier height vs scaled free energy minimum centre-of-mass 
separation for polymers in a tube of finite length. For the ring polymer systems,
the plot is $\Delta F/ND^{-1.90}$ vs $\lambda_{\min}/ND^{-0.67}$. For the linear
polymer system, 2.0$\Delta F/ND^{-1.90}$ vs 0.582$\lambda_{\min}/ND^{-0.67}$ is
shown. Results for different polymer lengths and tube diameters are shown. In each
data set for fixed $D$ and $N$, every data point corresponds to a different value of 
the tube length $L$. The inset shows the unscaled data.}
\label{fig:delF_scale_ring_cap}
\end{figure}

We now consider the scaling of the free energy barrier height, 
$\Delta F \equiv F(\lambda=0)-F(\lambda_{\rm min})$ with $N$, $L$ and $D$.
Results for ring polymer systems for different $D$ and $N$ are shown in
the inset of Fig.~\ref{fig:delF_scale_ring_cap}. Each data set for fixed
$N$ and $D$ corresponds to different values of $L$. To understand the scaling
properties of $\Delta F$, we use the same approach as for the analysis for
the $L$=$\infty$ system. First, employing Eq.~(\ref{eq:FringII}) for $\lambda=0$,
we note $F^{\rm (ring)}(0) = 2^{1+1/\nu} ND^{-1/\nu} u(0)$. This equation is expected
to be valid as long as the polymer extension length at complete overlap (i.e. 
$\lambda=0$) is less than full length of the tube, i.e. $L_{\rm ext}\leq L+D$.
At $\lambda=\lambda_{\rm min}$, the polymers are in regime~III. Consequently,
we employ Eq.~(\ref{eq:FringIII}), i.e.
\begin{eqnarray*}
F^{\rm (ring)}(\lambda_{\rm min})
= 2^{1/2\nu} ND^{-1/\nu} w\left(\frac{\lambda_{\rm min}}{2^{-3/2+1/2\nu} ND^{1-1/\nu}}\right)
\end{eqnarray*}
It follows that the free energy barrier height for the ring polymer system scales as 
\begin{eqnarray}
\Delta F^{\rm (ring)}(N,D,L) = ND^{-\alpha} 
h\left(\frac{\lambda_{\rm min}}{2^{-3/2+1/2\nu} ND^{-\beta}}\right) 
\label{eq:delFlmin}
\end{eqnarray}
where $h(x) \equiv 2^{1+1/\nu} u(0) - 2^{1/2\nu} w(x)$ and where the scaling
exponents are once again predicted to be $\alpha=1/\nu\approx 1.70$ and
$\beta=1/\nu-1\approx 0.70$. It follows that plotting results for $\Delta F/N^{-\alpha}$
vs $\lambda_{\rm min}/ND^{-\beta}$, the data should all collapse onto a universal curve.
Finally, using a similar approach to calculate the barrier height for linear
polymers, it is easily verified that the relationship to $\Delta F$ for ring polymers
follows the same scaling as for the $L$=$\infty$ free energy functions of 
Eq.~(\ref{eq:Fringlin}), i.e.
\begin{eqnarray}
\Delta F^{\rm (ring)}(\lambda_{\rm min};N,D) 
= 2^{1/2\nu} \Delta F^{\rm (lin)}(\lambda_{\rm min}/2^{-3/2+1/2\nu};N,D).
\label{eq:delFringlin}
\end{eqnarray}
Thus, for a given $D$ and $L$ (which determine $\lambda_{\rm min}$ and $N$), the function 
$\Delta F^{\rm (ring)}(\lambda_{\rm min})$ is related to $F^{\rm (lin)}(\lambda_{\rm min})$ 
by a scaling of $2^{1/2\nu}\approx 1.80$ along $F$ and a scaling of 
$2^{-3/2+1/2\nu}\approx 0.637$ along $\lambda$. 

In the main part of Fig.~\ref{fig:delF_scale_ring_cap}, we plot the scaled data using the 
values of the exponents measured earlier, i.e. $\alpha$=1.90 and $\beta$=0.67. The data 
for the linear polymers is also scaled by the same amounts as used in 
Fig.~\ref{fig:F.ring.linear.N200} to obtain data collapse for linear and ring polymer 
$F(\lambda)$ for $L$=$\infty$, i.e. a factor of 2.0 along $\Delta F$ and 0.582 along 
$\lambda_{\rm min}$.  Data collapse is good for the scaling predicted for $N$ and 
for the relationship between linear and ring polymers. However, it is somewhat poorer 
for the scaling with respect to $D$. Evidently, problems that lead to the discrepancy 
in the scaling with respect to $D$ in the $L$=$\infty$ results is further amplified in 
this scaling calculation of $\Delta F$ for finite $L$.

\begin{figure}[!ht]
\begin{center}
\includegraphics[width=0.4\textwidth]{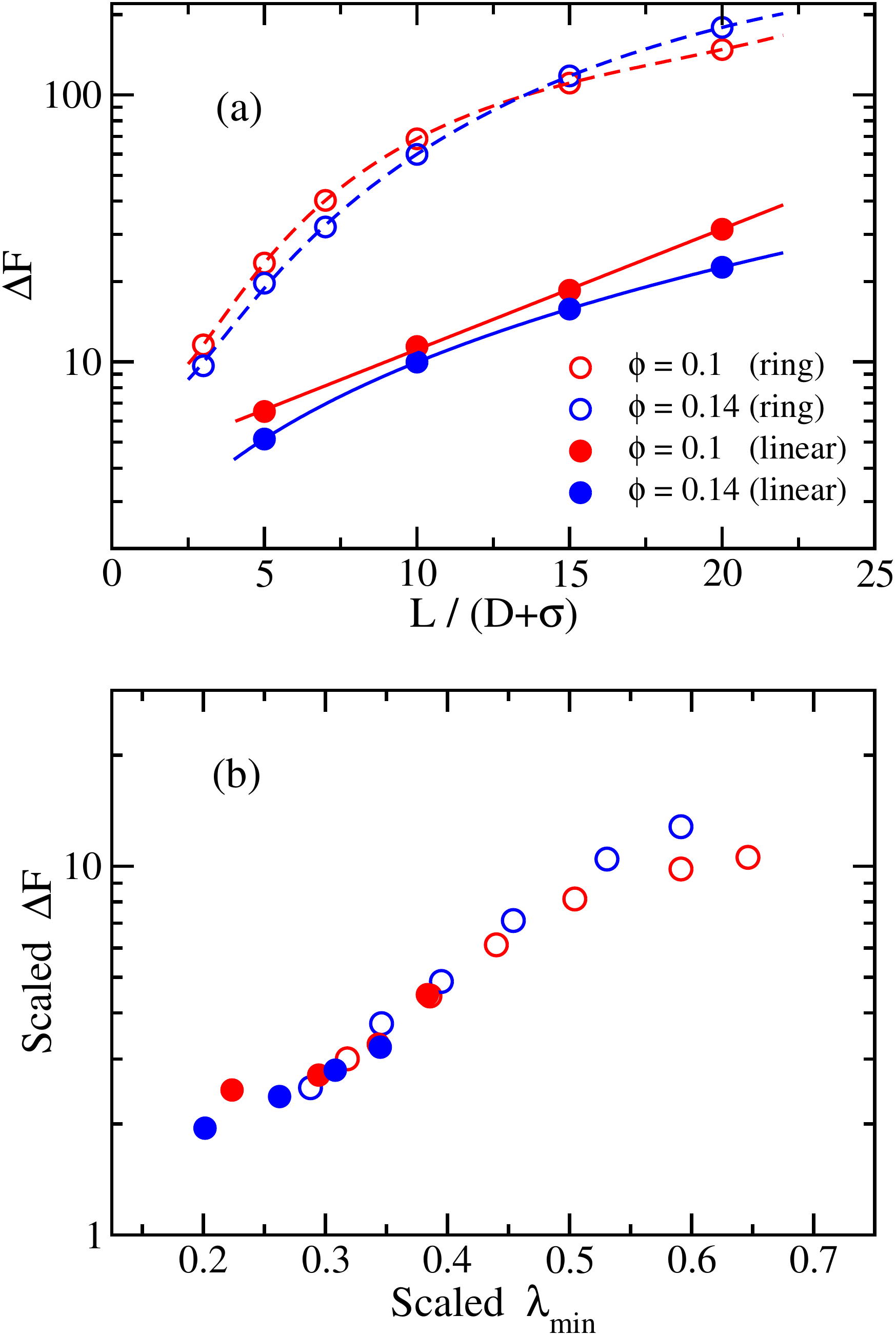}
\end{center}
\caption{(a) Free energy barrier height $\Delta F$ vs confinement tube aspect ratio $L/D$ 
for linear polymers (solid symbols) and ring polymers (open symbols). Simulations
were carried out for polymers of length $N$=200. The solid and dashed curves are guides 
for the eye. (b) Scaled data from (a). For the ring polymer systems,
the plot is $\Delta F/ND^{-1.90}$ vs $\lambda_{\min}/ND^{-0.67}$. For the linear
polymer system, 2.0$\Delta F/ND^{-1.90}$ vs 0.582$\lambda_{\min}/ND^{-0.67}$ is
shown. }
\label{fig:delF.DLratio}
\end{figure}

Next, we consider the effects of independently varying the confinement tube
aspect ratio and the packing fraction of the polymers. We define the aspect
ratio as $L/(D+\sigma)$, where $D+\sigma$ is the true diameter of the cylinder.
(Recall that $D$ is the diameter of the cylindrical volume accessible to
the monomer centres.) In addition, the packing fraction is given by
$\phi = 2Nv/V$,
where $v\equiv \pi \sigma^3/6$ is the monomer volume and where 
$V=\pi L (D+\sigma)^2/4 + \pi (D+\sigma)^3/6$ is the volume of the confining
tube. The free energy barrier height $\Delta F$ for ring and linear polymer 
systems for two different packing fractions and for several values of $L/(D+\sigma)$
are plotted in Fig.~\ref{fig:delF.DLratio}(a). Results for linear polymers with other 
$\phi$ values were presented in our earlier study.\cite{polson2014polymer} For both 
systems, the overlap free energy barrier increases monotonically with increasing 
confinement aspect ratio $L/(D+\sigma)$ for fixed monomer density. In the case of 
linear polymers, there is a slight decrease in $\Delta F$ with increasing density. 
As we noted previously,\cite{polson2014polymer} these trends are consistent with
the observation by Jung {\it et al.}\cite{jung2012ring} that polymer miscibility
decreases with increasing $L/(D+\sigma)$ and decreasing $\phi$. In the case of ring
polymers, there is no clear trend with varying $\phi$. At low $L/(D+\sigma)$ a lower
$\Delta F$ is observed for higher $\phi$, but the trend appears to reverse
at sufficiently high $L/(D+\sigma)$. Finally, we note that for any value of $L/(D+\sigma)$
and $\phi$, the overlap barrier height is significantly higher for ring
polymers than for linear polymers. 

It follows from Eqs.~(\ref{eq:delFlmin}) and (\ref{eq:delFringlin}) that the data
should collapse to a universal curve when plotting
$\Delta F/ND^{-\alpha}$ vs $\lambda_{\min}/ND^{-\beta}$ for the ring polymer data
and $2^{1/2\nu} \Delta F/ND^{-\alpha}$ vs $\lambda_{\min}/2^{1/2\nu} ND^{-\beta}$
for linear polymers, where $\alpha=1/\nu$ and $\beta=1-1/\nu$.
Figure~\ref{fig:delF.DLratio}(b) shows the data from
Figure~\ref{fig:delF.DLratio}(a) scaled such that the plot for ring polymers is
$\Delta F/ND^{-1.90}$ vs $\lambda_{\min}/ND^{-0.67}$ and that for linear polymers is
2.0$\Delta F/ND^{-1.90}$ vs 0.582$\lambda_{\min}/ND^{-0.67}$. (Thus, we use
the measured values of $\alpha$ and $\beta$ and the scaling factors for the relation
between the linear and ring polymer systems that gave the best data collapse for the
previous results.) The data collapse is good for the mid-range of scaled $\lambda_{\rm min}$.
At high values, the divergence of the two data sets for ring polymers likely
arises from the fact that $D$ becomes very small here, and the finite-size effects
associated with the blob model are amplified. At very low values of scaled $\lambda_{\rm min}$,
the deviation of the results for linear polymers is due to the fact that the tube
length becomes very short. Thus, the polymer extension length at full overlap ($\lambda$=0)
is affected by the longitudinal confinement, while the validity of Eq.~(\ref{eq:delFringlin})
rests on this assumption that this not be true. Despite these deviations, the scaling
arguments generally clearly do provide an explanation for the relationship between
the data for ring and linear polymers.

\begin{figure}[!ht]
\begin{center}
\includegraphics[width=0.4\textwidth]{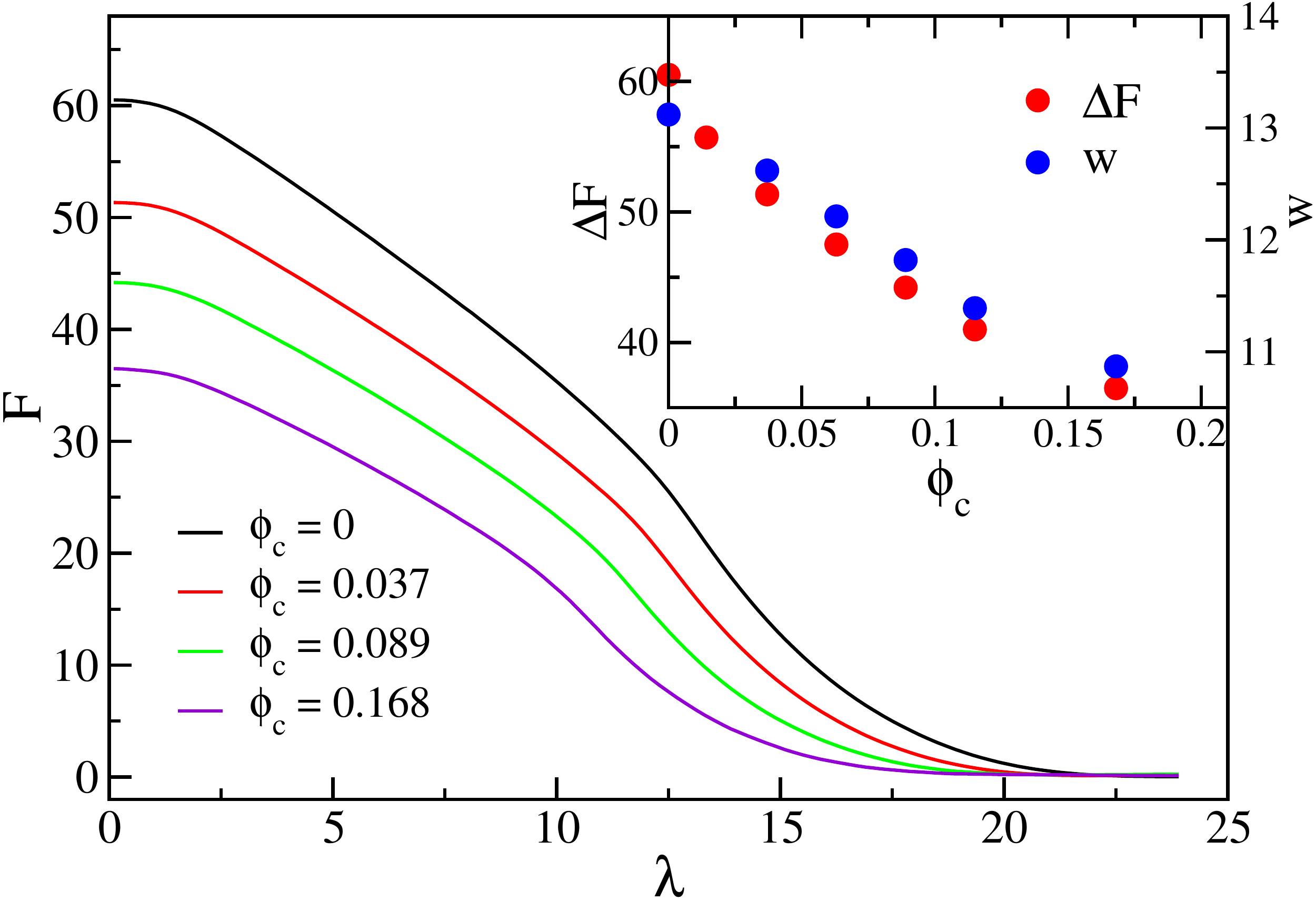}
\end{center}
\caption{Free energy vs $\lambda$ for ring polymers in a finite-length tube in
the presence of crowding agents. The tube has periodic boundary conditions at the ends.
Calculations were carried out for polymers of length $N$=80, in a tube with dimensions 
$L$=102 and $D=4$ and with crowders of diameter $\sigma_{\rm c}=1$. Results for several 
values of crowder packing fraction $\phi_{\rm c}$ are shown. The inset shows the variation
of the height $\Delta F$ and width $w$ of the free energy barrier with $\phi_{\rm c}$.  
The barrier width $w$ is determined by the inflection points in $F(\lambda)$.  }
\label{fig:F.N80.R2.5}
\end{figure}

Next we consider the effect of mobile crowding agents on the free energy functions.
We first examine the case of long confining cylinders with periodic boundary conditions.
This mimics the case of $L$=$\infty$ in the previous calculations.
In addition, the crowder diameter is set to be that of the monomers, i.e. $\sigma_{\rm c}$=1.
Figure~\ref{fig:F.N80.R2.5} shows free energies for a $N$=80 ring polymer system
for several values of the crowder packing fraction, $\phi_{\rm c}$. 
The inset shows the variation of the barrier height with $\phi_{\rm c}$. The free energy
decreases monotonically with increasing $\phi_{\rm c}$. This result is consistent with
the observation of Shin {\it et al.} for a comparable polymer system.\cite{shin2014mixing}
The inset also shows that the barrier width $w$ (arbitrarily defined to be the value of
$\lambda$ at the inflection point of $F(\lambda)$) also decreases with $\phi_{\rm c}$.

\begin{figure}[!ht]
\begin{center}
\includegraphics[width=0.45\textwidth]{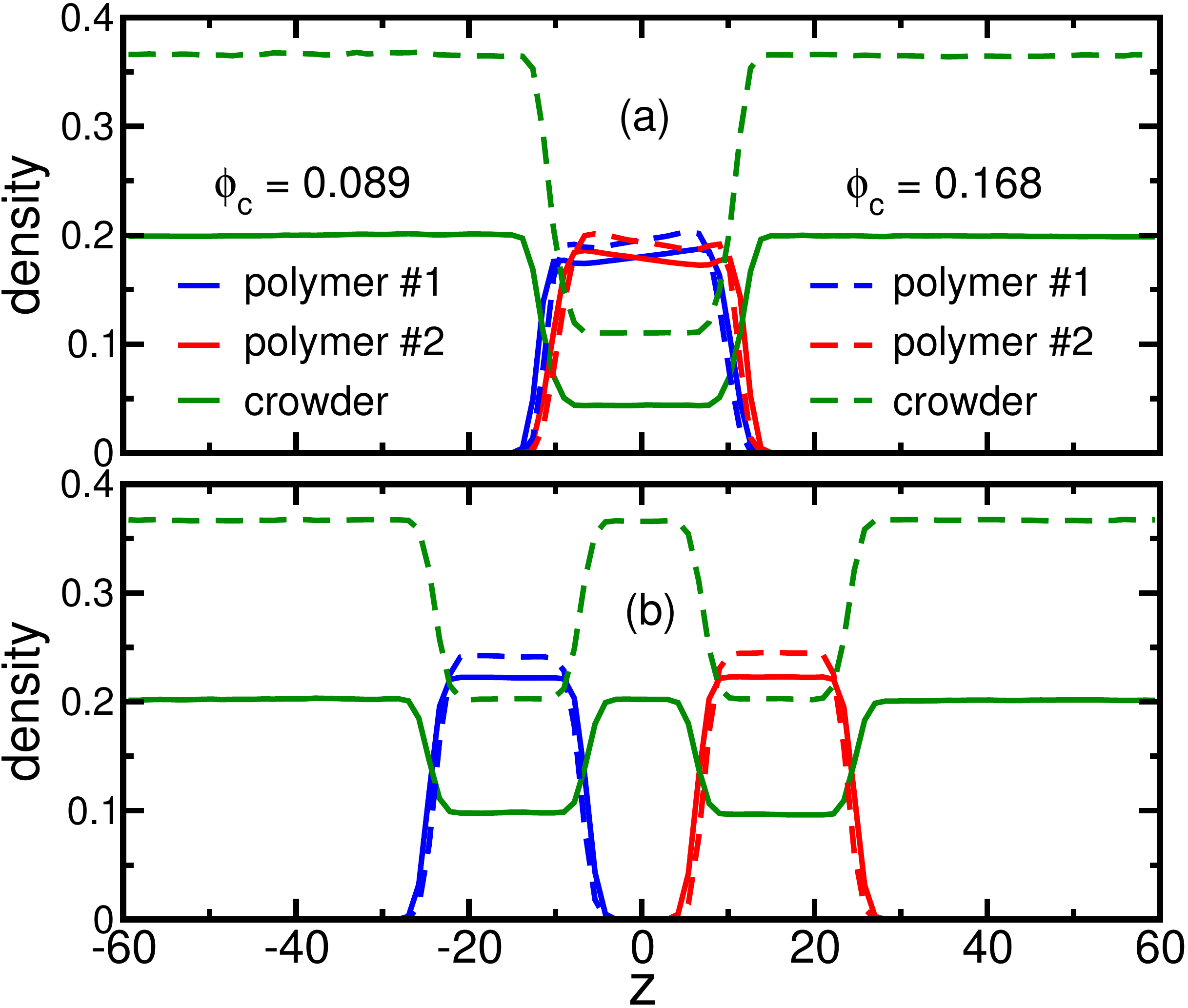}
\end{center}
\caption{Monomer and crowder density vs distance along the confining tube $z$.
Calculations were carried out for $N$=80, confinement dimensions $D$=4 and $L$=120,
and crowder diameter of $\sigma_{\rm c}$=1. Blue and red curves show the monomer 
densities for each polymer, and the green curves shows the crowder density. Solid curves 
are results for crowder packing fraction $\phi_{\rm c}$=0.089 and the dashed curves are 
results for $\phi_{\rm c}$=0.168. (a) Polymer separation is constrained to the range
$\lambda \in [0,2]$. (b) Polymer separation is constrained to the range $\lambda \in [29,31]$.
}
\label{fig:rhodist.N80.D4.rho_compare}
\end{figure}

Figure~\ref{fig:rhodist.N80.D4.rho_compare} shows the distributions of the crowding 
agents and the monomers of each polymer for two different values of $\phi_{\rm c}$. The top 
graph shows results for two polymers with nearly overlapping centres of mass,
and the lower graph shows results for the case for larger $\lambda$ where the polymers are 
well separated. As expected, the crowders have a reduced density in the region occupied by the 
polymers.  In addition the crowders are more effectively excluded from this region when the polymers 
overlap.  For both overlapping and non-overlapping polymers, increasing crowder density induces a 
slight compaction of the polymer along the tube axis. This is similar to the compaction effect 
observed for hard-sphere model systems of a single unconfined polymer in the presence of monomeric 
crowders (i.e.  $\sigma_{\rm c}=\sigma$).\cite{escobedo1996chemical} 
This compaction explains the trend in Fig.~\ref{fig:F.N80.R2.5} of decreasing barrier width 
with increasing $\phi_{\rm c}$: as the polymers compress along $z$ with increasing $\phi_{\rm c}$, 
the centre-of-mass distance at which the polymers just make contact also decreases.
The decrease in the polymer overlap free energy $\Delta F$ with increasing $\phi_{\rm c}$ likely 
arises principally for the same reason that crowding agents reduce the size of unconfined
polymers, i.e., the increase in translational entropy of the crowding agents with a reduction
in the volume occupied by the polymers. 

\begin{figure}[!ht]
\begin{center}
\includegraphics[width=0.4\textwidth]{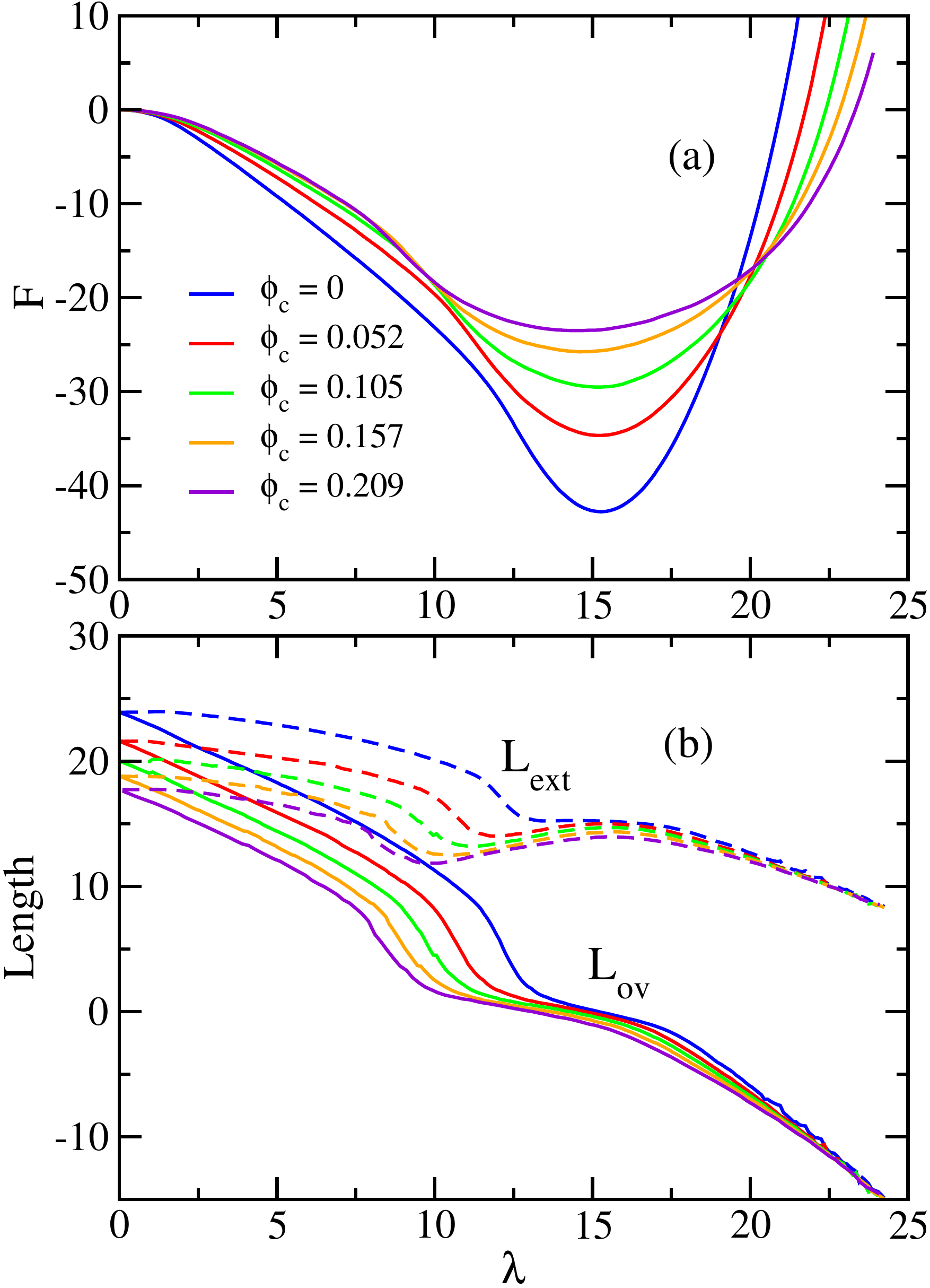}
\end{center}
\caption{(a) Free energy vs $\lambda$ for ring polymers confined in a tube of
finite length in the presence of crowding agents. Calculations were carried out
for polymers of length $N$=80 in a confining cylindrical tube of diameter $D$=4
and length $L$=28, with crowders of diameter $\sigma_{\rm c}=1$.  Results for several 
values of crowding agent packing fraction, $\phi_{\rm c}$, are shown. 
(b) Overlap length $L_{\rm ov}$ (solid curves) and extension length $L_{\rm ext}$ 
(dashed curves) vs $\lambda$ for the same systems as in (a).  }
\label{fig:F.lover.ring.cap.crowd.N80.R2.5.L28.sigc1.0}
\end{figure}

Next we consider the polymer/crowder system confined to tubes of finite length.
Figure~\ref{fig:F.lover.ring.cap.crowd.N80.R2.5.L28.sigc1.0}(a) shows free energy functions
for a system with of two ring polymers of length $N$=80 and crowders of size $\sigma_{\rm c}$=1
confined to a tube of length $L$=28. Results for several different crowder packing fractions
are shown.  As was the case for $L$=$\infty$,  $\Delta F\equiv F(0)-F(\lambda_{\rm min})$ 
decreases with increasing $\phi_{\rm c}$, again qualitatively consistent with the trend 
previously observed in Ref.~\citen{shin2014mixing}.
Figure~\ref{fig:F.lover.ring.cap.crowd.N80.R2.5.L28.sigc1.0}(b) 
shows the corresponding variation of the average $L_{\rm ov}$ and $L_{\rm ext}$ with $\lambda$.
Note that the polymer extension along the tube decreases with increasing $\phi_{\rm c}$
for all values of $\lambda$, i.e., the crowders have a small compactifying effect that
increases with density as was noted for $L$=$\infty$ above.

\begin{figure}[!ht]
\begin{center}
\includegraphics[width=0.4\textwidth]{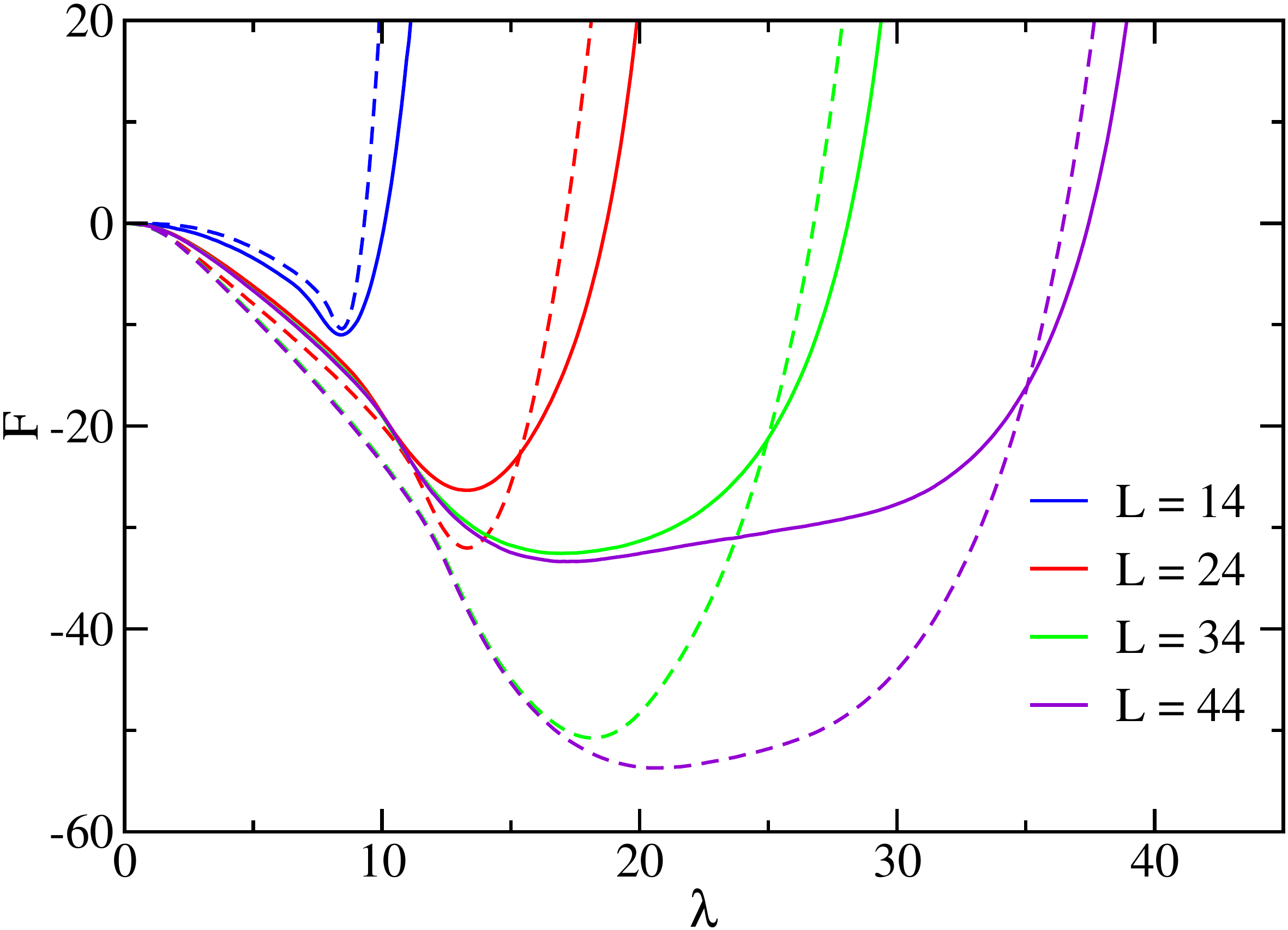}
\end{center}
\caption{Free energy vs $\lambda$ for ring polymers confined in a tube of
finite length in the presence of crowding agents. Calculations were carried out
for polymers of length $N$=80 in a confining cylindrical tube of diameter $D$=4
and with crowders of diameter $\sigma_{\rm c}$=1. Results are shown for crowder
packing fraction $\phi_{\rm c}$=0.105 (solid curves) and $\phi_{\rm c}$=0 (dashed curves)
for several values of confinement tube length, $L$.
}
\label{fig:F.ring.cap.crowd.N80.R2.5.phic0.105.sigc1.0}
\end{figure}

Figure~\ref{fig:F.ring.cap.crowd.N80.R2.5.phic0.105.sigc1.0} shows free energy functions
for tubes for several different tube lengths, each for $D=4$, $\phi_{\rm c}$=0 (i.e. no crowders)
and $\phi_{\rm c}$=0.105. For sufficiently long tubes, the presence of crowders
decreases the free energy barrier height, as in
Fig.~\ref{fig:F.lover.ring.cap.crowd.N80.R2.5.L28.sigc1.0}.
However, this effect on the barrier height diminishes as $L$ decreases, and at low 
$L$, crowding actually leads to a tiny increase in the barrier. To quantify this effect,
we define the ratio of the barrier heights with and without the crowders,
$r_{\Delta}\equiv \Delta F(\phi_{\rm c}=0.105)/\Delta F(\phi_{\rm c}=0)$. We find that
$r_{\Delta}$=0.63, 0.64, 0.82 and 1.06 for $L$=44, 34, 24 and 14, respectively.
In addition, for $L$=24, 34 and 44, $F$ decreases more rapidly with separation
distance $\lambda$ in the case of $\phi_{\rm c}$=0 than $\phi_{\rm c}$=0.105,
whereas the opposite opposite is true for the shortest tubes with $L$=14.
Thus, the effect of crowding on polymer segregation can be qualitatively different 
depending on the degree of longitudinal confinement.

\begin{figure}[!ht]
\begin{center}
\includegraphics[width=0.4\textwidth]{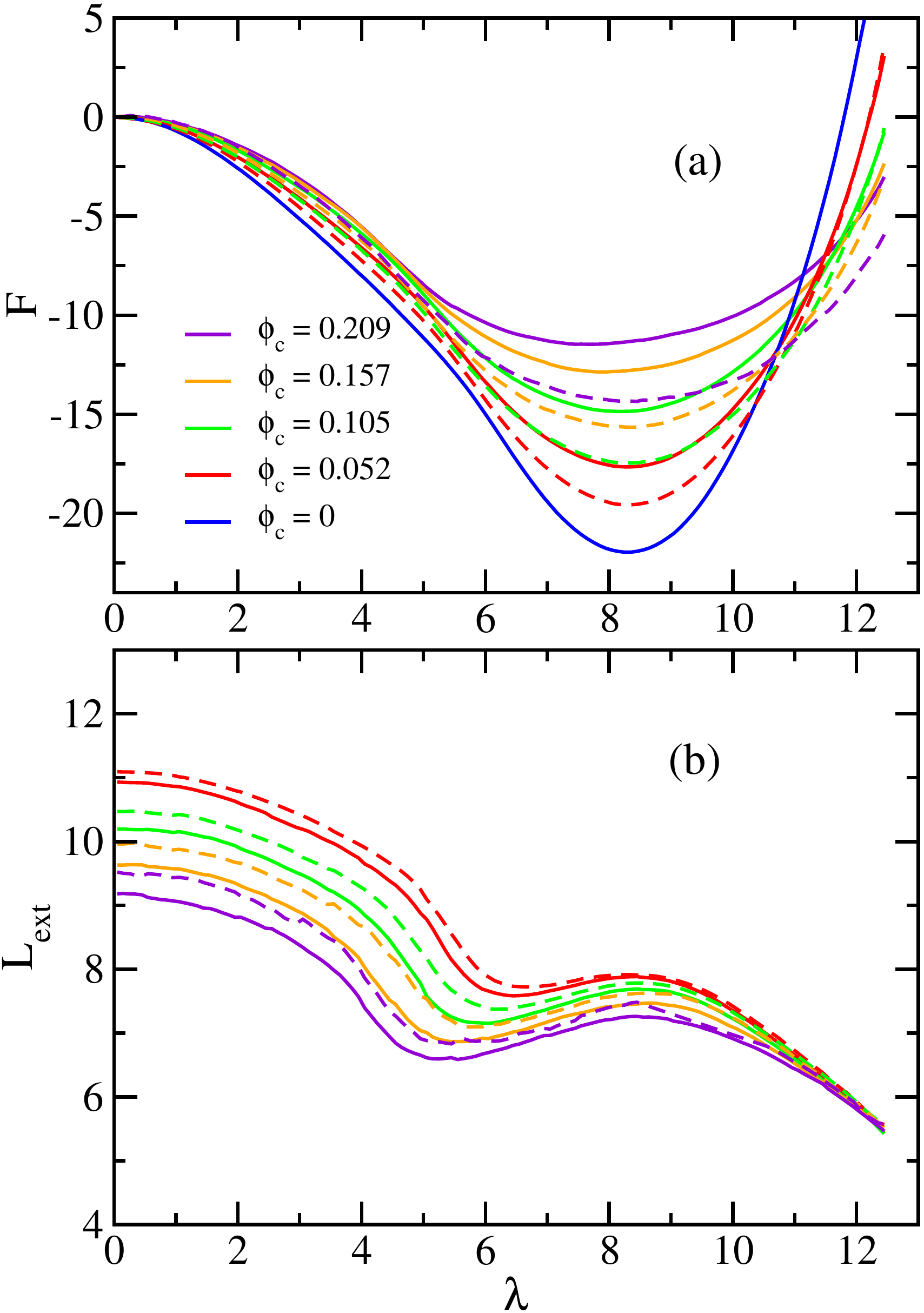}
\end{center}
\caption{(a) Free energy vs $\lambda$ for ring polymers confined in a tube of
finite length in the presence of crowding agents. Calculations were carried out
for polymers of length $N$=40 in a confining cylindrical tube of diameter $D$=4 
and length $L$=14.  Results are shown for crowders of diameter $\sigma_{\rm c}$=1
(solid curves) and $\sigma_{\rm c}$=0.5 (dashed curves) for several different
values of the crowder packing fraction $\phi_{\rm c}$. (b) Polymer extension length
vs $\lambda$ for the same calculations as in (a).
}
\label{fig:F.lover.ring.N40.R2.5.L14.sigc.compare}
\end{figure}

Finally, we briefly consider the effects of varying the size of the crowding agents. 
Figure~\ref{fig:F.lover.ring.N40.R2.5.L14.sigc.compare}(a) shows free energies for $N$=40 ring
polymers in a tube of dimensions $L$=14 and $D$=4. Results are shown for several different 
packing fractions, each for $\sigma_{\rm c}$=1 and $\sigma_{\rm c}$=0.5. At each fixed
$\phi_{\rm c}$, the effect of decreasing the crowder/monomer size ratio is to slightly
{\it increase} the free energy barrier. To understand this effect, we first note results
of previous studies on crowder size effects for linear polymers in infinite-length tubes.
For $\sigma_{\rm c} < \sigma$, decreasing $\sigma_{\rm c}$ at constant $\phi_{\rm c}$ 
has the effect of enhancing the depletion forces, which generally decreases the size of the 
polymer even to the point of inducing a collapse transition.\cite{kim2015polymer,jeon2016effects} 
By the arguments presented earlier, causing the polymers to become more compact
should {\it decrease} the overlap free energy. Note however in 
Fig.~\ref{fig:F.lover.ring.N40.R2.5.L14.sigc.compare}(b) that decreasing $\sigma_{\rm c}$
at fixed $\phi_{\rm c}$ leads to a slight {\it increase} in polymer extension length.

The origin of this increase may be related to an effect observed in recent studies
of a cylindrically confined linear\cite{kim2015polymer} and ring polymer\cite{jeon2017ring}
in the presence of crowders of size $\sigma_{\rm c}$=0.3. In those studies, it was noted 
that depletion forces between the polymer and cylinder wall can oppose the effects of 
monomer-monomer depletion forces. Provided the monomer-wall repulsion is not too 
strong, this leads to a regime in which the polymer extension increases with increasing 
$\phi_{\rm c}$.  In the case of a ring polymer, this tends to pull the two arms of the ring to 
opposite sides of the tube.\cite{jeon2017ring} In the present case, it is possible that 
such surface effects lead to the slightly greater extension observed for $\sigma_{\rm c}$=$0.5\sigma$ 
than for $\sigma_{\rm c}$=$\sigma$ for both overlapping and non-overlapping states.
Further complicating the picture, the polymer system used for 
Fig.~\ref{fig:F.lover.ring.N40.R2.5.L14.sigc.compare}
is subject to longitudinal confinement, unlike the case for those 
previous studies. The complex interplay of all of these features may combine to lead to
trends such as those observed here. A more thorough investigation is required to better 
clarify the effects of varying the crowder size but is beyond the scope of the
present work.

\section{Conclusions}
\label{sec:conclusions}

In this study we have used Monte Carlo simulations to investigate the segregation behaviour 
of two polymers under cylindrical confinement. We have measured the conformational free energy 
as a function of centre-of-mass distance between the polymers and examined the scaling 
of the free energy functions upon variation in the confinement dimensions, topology (i.e. ring
vs linear polymers) and crowder density and size. This work is a continuation of our previous study,%
\cite{polson2014polymer} in which we considered only linear polymers in the absence of
crowding. As in that study, we find that the free energy typically scales in a manner that is
in semi-quantitative agreement with predictions using simple scaling theoretical arguments.
In the absence of crowding, the theoretical model uses a combination of the de~Gennes blob
theory together with an approximation that the conformational free energy of two completely
overlapping polymers in a tube of cross-sectional area $A$ is equal to that of two non-interacting 
polymers confined to separate tubes of cross-sectional area $A/2$.\cite{jung2012ring} 
While the predicted scaling 
of the free energy with respect to confinement tube diameter and length, polymer length and
topology is qualitatively consistent with the simulation results, we find significant
deviations in the values of the measured scaling exponents from the predicted values.
Given that the polymers used in the simulations are only ${\cal O}(10^2)$ monomers in length, 
finite-size effects are likely a partial cause of this discrepancy.\cite{kim2013elasticity} 
However, a test of the approximation of Ref.~\citen{jung2012ring} reveals that it leads
to quantitatively poor predictions for the overlap free energy. This likely also contributes
to the discrepancies between the predicted and measured values of the scaling exponents.
In addition, there is no obvious reason why the accuracy of this approximation should improve 
for larger systems and, consequently, the scaling predictions for the free energy functions
are not expected to be accurate in the limit of large system size.

The presence of crowding agents was generally found to decrease the overlap free energy,
in accord with one previous result.\cite{shin2014mixing} This was observed in cases of both 
infinite- and finite-length confinement tubes, with the exception of the limiting case of 
very short tubes. This effect appears to arise from the tendency for crowding to compress the
polymers along the tube axis: such polymer compression increases the crowder translational
entropy by an amount that more than offsets the loss in polymer conformational entropy.
Thus, the translational freedom of the crowders is likely to be greatest when the polymers overlap. 
Decreasing the crowder diameter from $\sigma_{\rm c}=\sigma$ to $\sigma_{\rm c}=\sigma/2$
at constant packing fraction results in a slight reduction in the overlap free energy and
an increase in the extension length. These results are somewhat surprising given previous
measurements showing that depletion forces between monomers are enhanced by such a reduction
in crowder size. This effect may be due to the simultaneous enhancement of depletion forces
between the polymer and the confining wall, which was shown can lead to a regime in which 
the polymer extension increases with $\phi_{\rm c}$.\cite{kim2015polymer,jeon2017ring} 

While our investigation of the scaling properties of the overlap free energy functions 
have provided a reasonably complete picture for the case of polymers in the absence of 
crowding, the examination of crowding effects presented here requires much further
study.  In particular, it will be useful to carry out a thorough characterization of the 
effect of varying the crowder size over a much wider range of parameter space than that
considered here. In addition, it will be useful to examine the effects of crowding 
agent polydispersity,\cite{kim2015polymer} a feature that is clearly relevant to biological 
cells.

 This work was supported by the Natural Sciences and Engineering Research Council of 
 Canada (NSERC). We are grateful to the Atlantic Computational Excellence Network (ACEnet),
 Westgrid and Compute Canada for use of their computational resources.  

\providecommand*{\mcitethebibliography}{\thebibliography}
\csname @ifundefined\endcsname{endmcitethebibliography}
{\let\endmcitethebibliography\endthebibliography}{}


\end{document}